\documentclass[structabstract]{aa}  
\usepackage{graphicx}
\usepackage{txfonts}
\usepackage{longtable}
\usepackage{natbib}
\bibpunct{(}{)}{;}{a}{}{,} % to follow the A&A style
\def\Gd {G31.41$+$0.31}

\def\wat {H$_{2}$O}
\def\meth {CH$_{3}$OH}
\def\kms {km s$^{-1}$}
\def\Vlsr {$V_{\rm LSR}$}
\def\Vsys {$V_{\rm sys}$}

\def\Jcal {J1834$-$0301}

\begin{document}
   \title{A Double-Jet System in the \Gd\ Hot Molecular Core.}

%   \subtitle{I. Overviewing the $\kappa$-mechanism}

   \author{L. Moscadelli
          \inst{1}
          \and
          J.J. Li
          \inst{2}
          \and
          R. Cesaroni
          \inst{1}
          \and
          A. Sanna
          \inst{3}
          \and
          Y. Xu
          \inst{2}
          \and
          Q. Zhang
          \inst{4}
          }

   \institute{INAF-Osservatorio Astrofisico di Arcetri, Largo E.
Fermi 5, 50125 Firenze, Italy \\
              \email{mosca@arcetri.astro.it, cesa@arcetri.astro.it}
         \and
             Purple Mountain Observatory, Chinese Academy of
Sciences, Nanjing 210008, China \\
             \email{jjli@pmo.ac.cn, xuye@pmo.ac.cn}
%             \thanks{The university of heaven temporarily does not
%                     accept e-mails}
         \and
              Max-Planck-Institut f\"{u}r Radioastronomie,
Auf dem H\"{u}gel 69, 53121 Bonn, Germany \\
                \email{asanna@mpifr-bonn.mpg.de}
          \and
             Harvard-Smithsonian Center for Astrophysics, 60 Garden Street,
             Cambridge, MA 02138, USA \\
             \email{qzhang@cfa.harvard.edu}
             }
             
\offprints{L. Moscadelli, \email{mosca@arcetri.astro.it}}
\date{Received date; accepted date}

%   \date{Received September 15, 1996; accepted March 16, 1997}

% \abstract{}{}{}{}{} 
% 5 {} token are mandatory
 
  \abstract
  % context heading (optional)
  % {} leave it empty if necessary  
   {Many aspects of massive star ($\ga$10 $M_{\odot}$) formation are still unclear.
   In particular, 
%   to assess the role of circumstellar disks for accreting mass onto the
%   ``Massive Young Stellar Object'' (MYSO), and 
   the outflow properties  
   at close distance (100--1000~AU) from a ``Massive Young Stellar Object'' (MYSO) are not yet well established.}
%   Many aspects of massive star ($\ga$10 $M_{\odot}$) formation are still unclear.
%   In particular, we need to assess the role of circumstellar disks for accreting mass onto the
%   ``Massive Young Stellar Object'' (MYSO), and to estabilish the properties of the outflows 
%   at close distance (100--1000~AU) from the MYSO.}
  % aims heading (mandatory)
   {This work presents a detailed study of the gas kinematics towards the ``Hot Molecular Core'' (HMC) \Gd.}
%    This source has been previously intensively observed in thermal continuum and molecular line emissions
%    with several interferometers (VLA, PdBI, SMA), 
%    but, mainly owing to the limited angular resolution ($\ga$ a few 0\farcs1),
%    these observations were unable to provide a clear insight of the gas motion inside the HMC.}  
  % methods heading (mandatory)
   {To study the HMC 3-D kinematics at milli-arcsecond angular resolution,
   we have performed multi-epoch VLBI observations of the H$_2$O 22~GHz and CH$_3$OH 6.7~GHz
masers, and single-epoch VLBI of the OH 1.6~GHz masers.}
  % results heading (mandatory)
   {
%Distributed over the HMC, we detect 173 H$_2$O, 85 CH$_3$OH, and 69 OH maser features;
%proper motions are measured for 70 H$_2$O and 40 CH$_3$OH, persistent features.
%Water masers move significantly faster than methanol masers.
%Water maser, relative proper motions have a mean value of
%\ 34.0~\kms\ (with a mean error of \ 7.0~\kms), while for methanol masers
%the mean proper motion amplitude is \ 13.0~\kms\ (with a mean error of \ 4.0~\kms). 
Water masers present a symmetric spatial distribution with respect to the HMC center,
where two nearby (0\farcs2 apart), compact, VLA sources (labeled ``A'' and ``B'')
are previously detected.
The spatial distribution of a first group of water masers, named ``J1'',  
is well fit with an elliptical profile, and the maser proper motions mainly diverge
from the ellipse center, with average speed of 36~\kms. 
These findings strongly suggest that the ``J1'' water maser group traces
the heads of a young (dynamical time of \ $1.3 \, 10^3$~yr), powerful (momentum rate of \ $\simeq0.2$~$M_{\odot}$~yr$^{-1}$~km~s$^{-1}$), collimated (semi-opening angle $\simeq 10$\degr)
jet emerging from a MYSO located close (within $\approx$0\farcs15) to the VLA source ``B''.
Most of the water features not belonging to ``J1'' present an elongated ($\approx$2\arcsec\ in size), 
NE--SW oriented (PA$\approx$70\degr), S-shape distribution, which we denote with the label ``J2''.
The elongated distribution of the ``J2'' group and the direction of motion, approximately parallel to the direction of elongation, of most ``J2'' water masers suggests the presence of another collimated outflow, emitted from
a MYSO placed near the VLA source ``A''.
The proper motions of the CH$_3$OH 6.7~GHz masers, mostly diverging from the HMC center,
%and the regular variation of the OH 1.6~GHz maser LSR velocity (\Vlsr) along a direction approximately parallel to the ``J2'' major axis, 
also witness the expansion of the HMC gas driven by the ``J1'' and ``J2'' jets.
The orientation (PA$\approx$70\degr) of the ``J2'' jet agrees well with that (PA = 68\degr) of the
well-defined \Vlsr\ gradient across the HMC revealed 
%in several transitions of CH$_3$CN and also in the $^{12}$CO(2-1) line 
by previous interferometric, thermal line observations. Furthermore, the ``J2'' jet is powerful enough to sustain
the large momentum rate, 0.3~$M_{\odot}$~yr$^{-1}$~km~s$^{-1}$, estimated from the interferometric,
molecular line data in the assumption that
the \Vlsr\ gradient represents a collimated outflow. These two facts
lead us to favour the interpretation of the  \Vlsr\ gradient across the \Gd\ HMC in terms
of a compact and collimated outflow. 
}
  % conclusions heading (optional), leave it empty if necessary 
{}
%This work on the \Gd\ HMC confirms that VLBI maser, 3D kinematics can be relevant for the study
%of massive star-formation, this technique being presently the only one able 
%to resolve the typical sizes (several 100~AU) of massive stellar clusters, 
%and to study the proc(crossed by the ``J2'' maser strip)ess of mass accretion and ejection around a
%single MYSO.}

\keywords{ISM: individual objects (\Gd) -- ISM: kinematics and
dynamics -- masers -- techniques: interferometric}

   \maketitle
%
%________________________________________________________________

\section{Introduction}

    Understanding the formation of massive (OB-type) stars involves several theoretical and observational issues.
    From a theoretical point of view, models of massive star-formation are complicated
    by the need to consider the effects of the
    intense stellar radiation of a 
%``Massive Young Stellar Object'' (MYSO)
    MYSO, which, by heating, ionizing and 
    exerting pressure on
    the circumstellar gas, strongly influences the process of mass accretion.
    For a spherically symmetric infall, the strong radiation pressure may halt and even reverse the accreting flow
    and thus stop further growth of the stellar mass \citep{Yor02b}.
    Recent three-dimensional radiation-hydrodynamic models \citep{Kru09,Kui10,Kui11,Cun11} indicate that even 
    the most massive stars
    can be formed via accretion through a circumstellar disk, since the radiation
    pressure is decreased by conveying stellar photons along the disk axis and the ram pressure
    of the accreting gas increased by concentrating the accretion flow through the small
    disk solid angle. However, magnetohydrodynamic calculations \citep{Sei11} suggest 
    that the formation of a Keplerian disk can be suppressed by strong magnetic fields,
    since magnetic braking can efficiently remove the excess angular momentum of the accreting
    gas, lowering the centrifugal support against gravity. Another class of models (``competitive accretion'', \citet{Bon06}),      
    based on the observational evidence that massive stars are mainly found at the center of stellar clusters, 
    predict that massive protostars accrete not only the gas of the surrounding core but also more diffuse gas from the whole    
    molecular clump harbouring the cluster. According to this theory, accretion onto the massive protostar is 
    driven by the gravitational potential well which funnels the gas down to the cluster center. In this scenario,
     circumstellar disks might be   
    transient features prone to fragmentation and dispersion owing to strong interactions with cluster members.
     
    The most advanced simulations of massive star formation model both the accretion and the ejection 
    of mass self-consistently. The calculations by \citet{Cun11} are the first to consider also the feedback effects of 
    protostellar outflows, as well as protostellar radiative heating and radiation pressure on the infalling, dusty gas.    
    Previous works have considered how ionization and strong radiation pressure can
    affect the mass ejection, suggesting that the resulting increase in the plasma pressure and turbulence
    could contribute to the decollimation of molecular outflows from massive OB-type protostars \citep{Kon99,Yor02a,Vai11}.
    Since massive stars very often belong to multiple stellar systems, strong member interactions
    could in principle influence the outflow properties. In particular
    the tidal interaction between the disk and a companion star in a non-coplanar orbit 
    has been proposed as a viable mechanism for jet precession \citep{Ter99,Bat00}. 

    From an observational point of view, the two main difficulties to face 
    for a study of massive star formation
    are \ 1)~the relatively large distance (typically several kiloparsec) of the
    targets and \ 2)~the large extinction at visible and (often) near-infrared wavelengths.
    The second point precludes direct observation of the
    (proto)star, whose properties can be inferred only through the study
    of the radio to mid-infrared radiation from the circumstellar gas reprocessing the
    stellar photons. The first problem requires angular resolutions  \ $\le$0\farcs1
    to observe the scales (less than a few hundreds AU) relevant to the study of  
    mass accretion and ejection.  Prior to the advent of the Atacama Large Millimeter Array (ALMA),
    the angular resolution ($\ge$ a few 0\farcs1) of millimeter interferometers (PdBI, SMA) was in most cases inadequate 
    both to resolve massive accretion disks (with model-predicted sizes of a few hundreds AU)
    and to identify the protostellar outflow emitted from a ``single'' MYSO system (hidden in the complex emission pattern
    resulting from the interaction of the cluster outflows).
    
    The limited angular resolution can explain the failure to detect disks in O-type stars, and the
    small number of bona-fide detections obtained for B-type stars \citep{Ces07}.
    In association with the most luminous MYSOs one finds only huge ($\la$0.1~pc), massive (a few 100~$M_{\odot}$)
    cores, with velocity gradients suggesting rotation. These objects, named ``toroids'',
    are likely non-equilibrium structures, 
    because the ratio between the accretion time scale and 
    the rotation period is much smaller in these toroids than in accretion disks \citep{Ces07,Bel11}.
    The general properties of molecular outflows in massive star-forming regions 
    are presently known mainly through single-dish surveys, 
    with $\ga$10\arcsec angular resolution \citep{Beu02b,Wu05,Zha05,Lop09},
    of typical outflow tracers (such as CO and its isotopomer $^{13}$CO). These studies have emphasized a 
    good correlation between the momentum rate of the outflow and the MYSO bolometric luminosity. 
    On the other hand, Very Large Array (VLA) sub-arcsecond observations have revealed a few thermal jets (HH~80-81 
    \citep{Mar98}; IRAS~16547$-$4247 \citep{Rod08};  IRAS~20126$+$4104  
    \citep{Hof07}) associated with intermediate-~and~high-mass (proto)stars. It is interesting that these
    (proto)stars, although being of moderate luminosity (10$^4$--10$^5$~$L_{\odot}$), 
    eject very powerful, collimated thermal jets and molecular outflows,
    whose momentum rates ($\sim$0.1~$M_{\odot}$~yr$^{-1}$~km~s$^{-1}$) are close 
    to the upper limit measured in the single-dish surveys of molecular outflows.
    These results call for high-angular resolution ($\la$0\farcs1) observations of a larger sample of outflows
from MYSOs to ascertain if the most compact and youngest flows (not observed in the single-dish surveys) 
are intrinsically more powerful and collimated. If this were confirmed, the outflow evolution 
for high-mass (proto)stars would proceed similarly to the low-mass case, where outflows from Class~0 objects 
are more intense and collimated than for Class~I sources \citep{Bon96}.  
 
   Since several years, we are conducting a campaign of multi-epoch Very Long Baseline Interferometry (VLBI)
    observations of the strongest interstellar molecular masers (OH at 1.6~GHz, CH$_3$OH at 6.7~GHz, and H$_2$O at 22~GHz) towards a small sample of massive star-forming regions \citep{Mos07,Mos11,San10a,God11,Li12}. Multi-epoch VLBI observations,
    achieving angular resolutions of \ 1-10~mas, 
    permit to measure the proper motions of the maser emission centers, and, knowing the source distance and the maser
%    LSR velocity (\Vlsr) 
    \Vlsr\ (from Doppler effect), one can reconstruct the 3-D~kinematics of the masing gas.
    Our results demonstrate the ability of maser emission to trace kinematic structures close to (within tens/hundreds AU from)   
    the MYSO, 
    revealing the presence of fast wide-angle and/or collimated outflows, traced by the H$_2$O masers, and 
    of rotation and infall, indicated by the CH$_3$OH masers. This work reports on our recent VLBI observations
    of the molecular masers associated with the 
%    ``Hot Molecular Core'' (HMC) 
     HMC \Gd, whose gas kinematics has been   
    investigated in the last years using the best available cm and mm interferometers (VLA, SMA, PdBI).
    
    \Gd\ is located at a kinematic distance of 7.9~kpc, and its high bolometric luminosity of \ 3$\times$10$^5$~$L_{\odot}$ \citep{Ces94} hints at the presence of embedded high-mass (proto)stars. The HMC in \Gd\ (at a systemic LSR velocity, \Vsys\ = 97.0~\kms) was originally imaged with the VLA in the high-excitation (4,4) inversion transition of ammonia by \citet{Ces94b}, and successively mapped by  \citet{Bel04} and \citet{Ces11}, using the PdBI and SMA interferometers, respectively, in the (12-11) rotational transitions of methyl cyanide (CH$_3$CN). These latter observations, achieving an angular resolution
 of 0\farcs8, reveal a well-defined velocity gradient across the HMC, oriented in the NE--SW direction. Considering the
 large size ($\approx$0.1~pc) and mass (several 100~$M_{\odot}$) of this structure, \citet{Bel04}~and~\citet{Ces11} 
 interpret the velocity gradient in terms of a toroid in pseudo-Keplerian rotation around a stellar cluster of
 $\approx$10$^3$~$M_{\odot}$, tightly packed at the center of the HMC. In contrast to this interpretation, 
 \citet{Ara08} explain the velocity gradient in terms of a compact outflow on the basis of 
  the NE--SW bipolar morphology of thermal methanol (at 44~GHz), 
 observed with the VLA at an angular resolution of $\approx$0\farcs5.
%  Since the HMC in \Gd\ is one of the most notable examples
% of bona-fide toroids, distinguishing between rotation and expansion in this source can help to shed light on the
% nature of other similar objects. 
In an attempt to decipher the gas kinematics inside the
 \Gd\ HMC, we have observed the 
 OH 1.6~GHz, CH$_3$OH 6.7~GHz, and H$_2$O 22~GHz masers associated with the core, using the VLBI.
 The observational details are given in Sect.~\ref{mas_obs}, while the results
 are illustrated in Sect.~\ref{res} and discussed in Sects.~\ref{jets_exp}~and~\ref{discu}.
 The conclusions are drawn in Sect.~\ref{conclu}.

% Acronims to define:  HMC    
% Terms to define: \Vsys
    
\section{VLBI Maser Observations and Data Reduction}

\label{mas_obs}

We conducted VLBI observations of the \wat\ 22~GHz and \meth\ 6.7~GHz masers (at
four epochs) and of the OH 1.665/1.667~GHz masers (at a single epoch) 
toward \Gd. 
Table~\ref{tbl:mas-abs-pos} lists observing epochs and beams, and reports the derived
maser positions.
In order to
determine the maser absolute positions, we performed
phase-referencing observations in fast switching mode between the
maser source and the calibrator \Jcal. This calibrator has an
angular offset from the maser source of 3\fdg8 and belongs to the
list of sources of the International Celestial Reference Frame 2
(ICRF2). Its absolute position is known with a precision better than
$\pm$1.0~mas, and its flux measured at S~and~X~bands
is 250 and 280~mJy~beam$^{-1}$, respectively. Four fringe finders,
J1642$+$3948, J1751$+$0939, J1800$+$3848~and~J2253$+$1608, were
observed for bandpass, single-band delay, and instrumental
phase-offset calibration.

Data were reduced with AIPS following the VLBI spectral line
procedures. For a description of the general data calibration and
the criteria used to identify individual masing clouds, derive their
parameters (position, intensity, flux and size), and measure their
(relative and absolute) proper motions, we refer to the recent paper
on VLBI observations of \wat\ and \meth\ masers by \citet{San10a}.
For each VLBI epoch, except for the fourth epoch on \meth\ masers (see Sect~\ref{obs_evn}),
 \wat\ and \meth\ maser absolute positions have
been derived by using two ways, direct phase-referencing and inverse
phase-referencing, and the two procedures always gave consistent
results. For the weak OH masers, we only use direct phase-referencing 
to determine their absolute positions, because inverse phase-referencing is
hindered by the insufficient number of maser (phase calibration) solutions. 
%
% Do inverse and direct phase-reference work fine for the OH masers, as well ?
%
%\meth\ and OH maser absolute positions have been derived
%only by using direct phase-referencing.
%
The derived \emph{absolute} proper motions have been corrected for the
apparent proper motion due to the Earth's orbit around the Sun
(parallax), the Solar Motion and the differential Galactic Rotation
between our LSR and that of the maser source. We have adopted a flat
Galaxy rotation curve ($R_0 = 8.3\pm0.23$~kpc, $\Theta_0 = 239\pm7
$~\kms) \citep{Bru11}, and the Solar motion ($U = 11.1
^{+0.69}_{-0.75}$, $V = 12.24 ^{+0.47}_{-0.47}$, and $W = 7.25
^{+0.37}_{-0.36}$~\kms) by \citet{Sch10}, who have recently revised 
the Hipparcos satellite results.
We antecipate that the absolute proper motions of water and methanol masers
are dominated by a large systematic effect due to
the uncertainties in the source distance and Solar and Galactic
motion, and/or by a peculiar velocity of the maser source with
respect to its LSR reference system. Being severely affected
by this systematic deviation, the absolute proper motions 
cannot be used to describe the maser kinematics
around the MYSO, to which purpose we employ the relative proper motions. 

\subsection{VLBA: 22.2~GHz \wat\ Masers}

We observed \Gd\ (tracking center: R.A.(J2000) = $18^{\rm
h}47^{\rm m}34.35^{\rm s}$ and Dec.(J2000) = $-01\degr12\arcmin
45.6\arcsec$) with the NRAO\footnote{The National Radio Astronomy Observatory is operated by
Associated Universities, Inc., under cooperative agreement with the National
Science Foundation.}
Very Long Baseline Array (VLBA) in the $6_{16}-5_{23}$ \wat\ transition
(rest frequency 22.235079~GHz). The observations (program code:
BM286) consisted of 4 epochs: 2008 November 23, 2009 January 18,
2009 March 18, and 2009 May 18. Following \citet{Rei99}, to reduce
systematic errors owing to correlator atmospheric mismodeling, we
observed three ``geodetic'' blocks, each lasting about 40~min, before
the start, in the middle, and after the end of our phase-reference
observations. The ``geodetic-type'' data were taken in left circular
polarization with eight 4~MHz bands spanning a bandwidth of 480~MHz.
The fast-switching phase-referenced observations consisted of two
$\approx$3~hour blocks, during which we recorded the dual circular
polarization through a 16~MHz bandwidth centered at 
\ $V_{\rm LSR}$ = 96~\kms. The data were processed with the VLBA
FX correlator in Socorro (New Mexico) using an averaging time of 1~s
and 1024 spectral channels.

Images were produced with natural weighting, cleaned and restored
with the naturally-weighted beam, having a Full Width at Half Maximum (FWHM)
size of about 1.6~mas $\times$ 1.0~mas at a position angle (PA) of
6\degr \ (East of North), showing small variations from epoch to epoch
(see Table~\ref{tbl:mas-abs-pos}). The interferometer
instantaneous field of view was limited to $\sim$2\farcs5. In each
observing epoch the on-source integration time was $\sim$2.5~h,
resulting in an effective rms noise level on the channel maps
(1$\sigma$) in the range 6--10~mJy~beam$^{-1}$. The spectral
resolution was 0.21~\kms.

\subsection{VLBA: 1.665/7~GHz OH Masers}

We observed \Gd\ (tracking center: R.A.(J2000) = $18^{\rm h}47^{\rm
m}34.3^{\rm s}$ and Dec.(J2000) = $-01\degr12\arcmin 46.0\arcsec$)
with the VLBA in the ``main'' lines (rest frequencies of
1.665401~GHz and 1.667359~GHz) of the hyperfine
splitting of the $^{2}\Pi_{3/2} J = 3/2$ OH level on September
26 2008 (program code: BM286E). During a run of $\sim$8~h, we alternated
the observations of either ``main'' lines, recording the dual
circular polarization with one bandwidth of 1~MHz (centered at the
maser $V_{\rm LSR}$ of 97.0~\kms) and three 4~MHz bandwidths. The
4~MHz bandwidths were used to increase the SNR of the weak L-band
signal of the continuum calibrator. The data were processed with the
VLBA FX correlator in two correlation passes using 1024 and 128
spectral channels for the 1~MHz and 4~MHz bands, respectively. In
each correlator pass, the data averaging time was 2~s.

Images were produced with natural weighting, cleaned and restored
with the naturally-weighted beam, having a FWHM size of 19~mas $\times$ 11~mas
at a PA of $-$3\degr, and 24~mas $\times$ 15~mas at a PA of 35\degr,
for the 1665 and 1667~MHz masers, respectively. The interferometer
instantaneous field of view was limited to 7\farcs2. The on-source
integration time was $\sim$1.2~h, resulting in an effective rms
noise level in each velocity channel of $\sim$17~mJy~beam$^{-1}$.
The spectral resolution attained across the maser 1~MHz band
 was 0.18~\kms.

\subsection{EVN: 6.7~GHz \meth\ Masers}
\label{obs_evn}

We observed \Gd\ (tracking center: R.A.(J2000) = $18^{\rm
h}47^{\rm m}34.33^{\rm s}$ and Dec.(J2000) = $-01\degr12\arcmin
47.0\arcsec$) with the European VLBI Network\footnote{The European VLBI Network is a joint facility of European,
Chinese and other radio astronomy institutes founded by their national
research councils.} (EVN) in the
$5_{1}-6_{0}A^{+}$ \meth\ transition (rest frequency 6.668519~GHz).
Data were taken at 4 epochs (program code: EM071)
separated by $\sim$1~year: 2009 March 15, 2010 March 11, 2011
March 2, and 2012 February 24. In the first two epochs, the antennae
involved in the observations were Cambridge, Jodrell2, Effelsberg, Onsala,
Medicina, Noto, Torun, Westerbork and Yebes. Because of technical problems,
the Noto antenna could not take part to the observations of the third and fourth
epochs. The Knockin antenna replaced the Cambridge antenna in the third epoch,
and the Cambridge antenna did not observe the fourth epoch. 
During a run of $\sim$8~h, we recorded the dual
circular polarization through eight adjacent bandwidths of 2~MHz, one of them
centered at the maser \Vlsr\ of 99.0~\kms. The eight 2~MHz bandwidths
were used to increase the SNR of the weak continuum calibrator. The
data were processed with the MKIV correlator at the Joint Institute
for VLBI in Europe (JIVE-Dwingeloo, the Netherlands) in two correlation passes,
using 1024 and 128 spectral channels to correlate the maser 2~MHz bandwidth
and the whole set of eight 2~MHz bandwidths, respectively. 
In each correlator pass, the data averaging time was 1~s.

In addiction to the
absence of Cambridge and Noto, the fourth observing epoch was also severely affected by other
problems: the Jodrell2 receiver was warm ($T_{sys} \approx 250$~K); Medicina
suffered a computer problem, after which the fringe amplitudes were much reduced; 
because of strong winds, Onsala was stowed half of the observing time.
The good data were not sufficient to reliably measure the maser absolute position,
which has been merely estimated from a linear fit of the positions of the first three epochs (see Table~\ref{tbl:mas-abs-pos}). 

Reflecting the changes in the antennae taking part in the observations,
the FWHM size of the naturally-weighted beam
varied over the four epochs
from about 8~mas $\times$ 5~mas (PA = -17\degr) to
about 12~mas $\times$ 3.5~mas (PA = -37\degr)
(see Table~\ref{tbl:mas-abs-pos}). 
At the first two epochs, the maser images have been cleaned and restored
using the naturally-weighted beam. 
At the third and fourth epoch, to prevent apparent changes 
in the maser structure owing 
to the variation of the naturally-weighted beam, 
we have restored the maser images 
using a beam, 8.3~mas $\times$ 4.8~mas at PA = $-17\degr$, average
of the naturally-weighted beams of the first and second epochs.
The interferometer
instantaneous field of view was limited to $\sim$33\arcsec.
Using an on-source integration time of $\sim$2.3~h, the
effective rms noise level on the channel maps (1$\sigma$) varied in
the range \ 5--13~mJy~beam$^{-1}$. The spectral resolution attained 
across the maser 2~MHz band was 0.09~\kms.

\begin{table*}
\caption{\label{tbl:mas-abs-pos} Maser positions, brightnesses and corresponding synthesized beams.}
\begin{tabular}{lccllclc}
\hline \hline
Maser & Feature & Epoch of & R.A.(J2000) & Dec.(J2000) & I$_{\rm peak}$ & Major $\times$ Minor, PA \\
 & Number & Observation & $\mathrm{(^h\;\;\;^m\;\;\;^s)}$ & $(\degr\;\;\;\arcmin\;\;\;\arcsec)$
  & (Jy beam$^{-1}$) & (mas $\times$ mas, deg) \\
\hline
H$_{2}$O & 1 & 2008 Nov 23 & 18~47~34.31259 & $-$01~12~46.6384  & 115 &  1.6 $\times$ 0.8 , $-$7  \\
         &   & 2009 Jan 18 & 18~47~34.31257 & $-$01~12~46.6394  & 144 &  1.6 $\times$ 1.1 , $-$1  \\
         &   & 2009 Mar 18 & 18~47~34.31256 & $-$01~12~46.6402  &  75 &  1.4 $\times$ 1.0 , 13 \\
         &   & 2009 May 18 & 18~47~34.31254 & $-$01~12~46.6412  &  40 &  1.8 $\times$ 1.1 , 20  \\
\hline
CH$_{3}$OH & 1 & 2009 Mar 15 & 18~47~34.28652 & $-$01~12~45.5777 & 8.3 & 7.8 $\times$ 5.0 , $-$17 \\
           &   & 2010 Mar 11 & 18~47~34.28636 & $-$01~12~45.5837 & 9.0 & 8.8 $\times$ 4.6 , $-$18 \\
           &   & 2011 Mar 2 & 18~47~34.28625 & $-$01~12~45.5879 & 10.9\tablefootmark{a} & 11.8 $\times$ 3.3 , $-$36   \\
           &   & 2012 Feb 24 & 18~47~34.28610 & $-$01~12~45.5934 & 10.1\tablefootmark{a} & 11.6 $\times$ 3.3 , $-$38 \\
\hline
OH-1665 & 1~LCP & 2008 Sep 26 & 18~47~34.27583 & $-$01~12~46.3282 & 4.0 & 19 $\times$ 11 , $-$3 & \\
OH-1667 & 1~LCP & 2008 Sep 26 & 18~47~34.27574 & $-$01~12~46.3287 & 5.0 & 24 $\times$ 15 , 35 & \\
\hline
\end{tabular}
\tablefoot{Column~1 indicates the maser species; column~2 gives
the label number (and, for the OH masers, polarization) of the phase-reference feature, as reported in Tables~\ref{oh1665_tab},
\ref{oh1667_tab}, \ref{h2o_tab}, and \ref{ch3oh_tab}; column~3 lists the observing date;
columns~4~and~5 report the absolute position in Equatorial coordinates; column~6 gives the
peak intensity; column~7 reports the (major and minor) FWHM size
and PA of the naturally-weighted beam. The PA of the beam is
defined  East of North. The accuracy in the derived maser absolute positions is \ $\sim$1~mas in both Equatorial coordinates.\\
\tablefoottext{a}{Peak intensity from the image restored using a beam 
with FWHM size of \ 8.3~mas $\times$ 4.8~mas at PA = $-17\degr$.}
}
\end{table*}

\section{Results}
\label{res}

\begin{figure*}
\centering
\includegraphics[width=14.5cm]{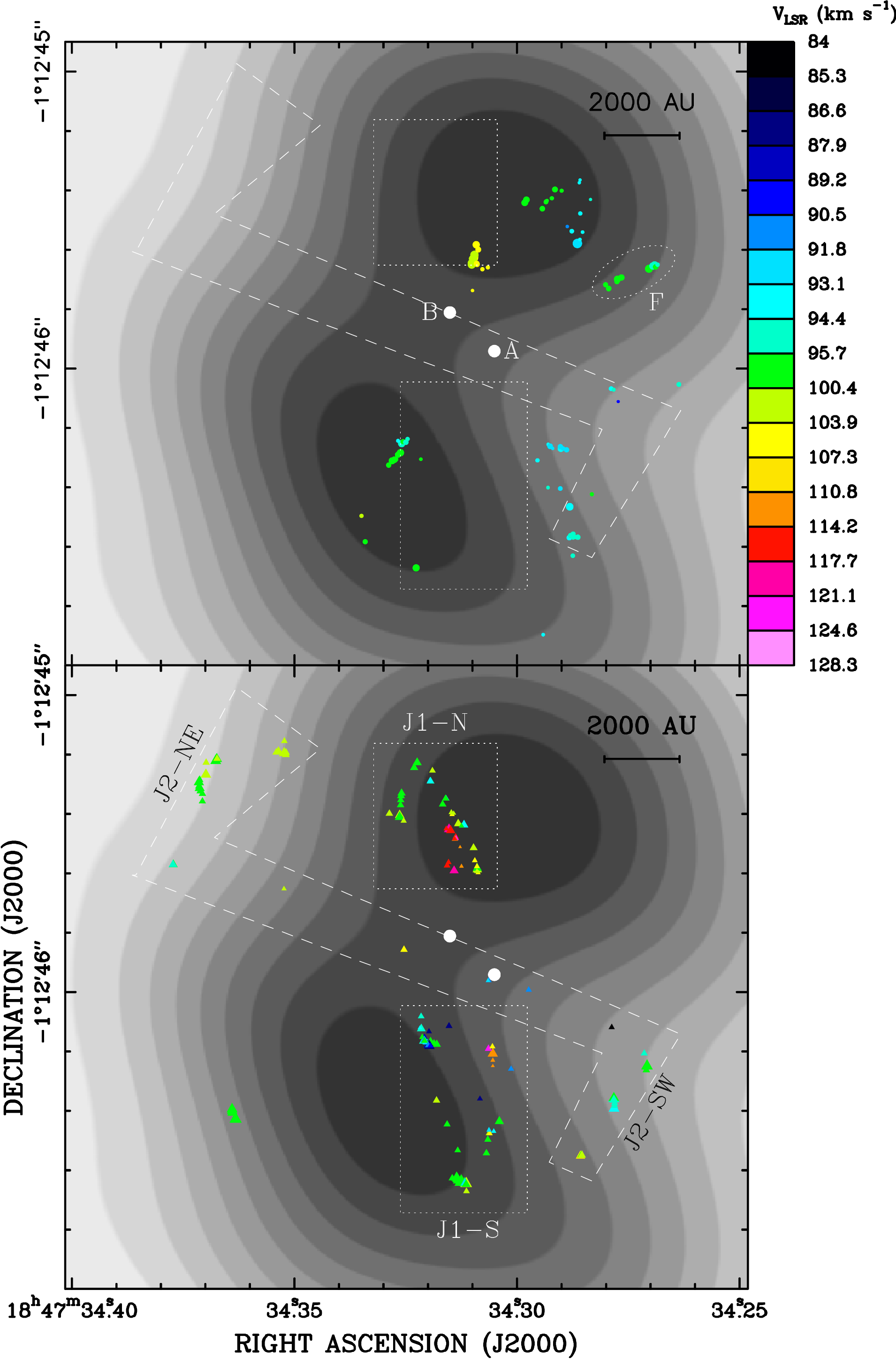}
\caption{\small {\it Lower} and {\it upper} panels show the absolute position 
and LSR velocity of the 22~GHz water ({\it filled triangles}) 
and 6.7~GHz methanol ({\it dots}) maser features, respectively.
Colors denote the maser \Vlsr\ according to the color-velocity conversion code
shown on the right-hand side of the upper panel. Symbol area is proportional to the
logarithm of the feature intensity. In both panels, the {\it grey-scale map} reproduces the
integrated emission of the CH$_{3}$CN~(12-11)~K=4 line (at $\approx$221~GHz) 
observed with the SMA (beam $\approx$0\farcs8) by \citet{Ces11}. Contour levels
are from \ 0.15~Jy~beam$^{-1}$ to \ 2.3~Jy~beam$^{-1}$ by steps of \ 0.27~Jy~beam$^{-1}$.
The {\it white big dots} denote the position of the two  
VLA compact sources, named ``A'' and ``B'', detected by \cite{Ces10}.
In both panels, the two {\it dotted rectangles} and the S-shape {\it dashed polygon}
encompass the area where the ``J1'' and ``J2'' water maser groups
distribute, respectively (see Sect~\ref{res_wat}). 
The northern and southern ``J1'' clusters are named ``J1-N'' and ``J1-S'', respectively.
The water maser clusters at the southwestern and northeastern ends of the ``J2'' distribution
are named ``J2-SW'' and ``J2-NE'', respectively.
In the upper panel, the {\it dotted ellipse} surrounds a cluster of 6.7~GHz methanol masers,
identified with the name ``F'',
more detached from the area of water maser detection.}
\label{mas_dis_1}
\end{figure*}

\subsection{22~GHz Water Masers}
\label{res_wat}

The lower panel of Fig.~\ref{mas_dis_1} shows absolute positions and LSR velocities
of the detected water maser features, whose parameters (derived by fitting a 2-D elliptical Gaussian
to the maser intensity distribution) are listed in Table~\ref{h2o_tab}.
Over four epochs, we detected 173 distinct maser features,
whose intensity ranges from the detection threshold of \ $\approx$50~mJy~beam$^{-1}$
\ up to $\approx$100~Jy~beam$^{-1}$, and with \Vlsr\ varying from 84~\kms\ to 128~\kms.
Interestingly, the distribution of the water masers is approximately symmetric
with respect to the position of the two nearby, compact VLA continuum sources ``A'' and ``B'', observed
by \citet{Ces10} roughly at the center of the \Gd\ HMC. At comparable offsets to the North and South
of the continuum sources, one observes two similar clusters of water masers 
(enclosed in the white dotted rectangles in Fig.~\ref{mas_dis_1}), both clusters having the shape
of acute bows, with the bow tip pointing to the North (South) for the northern (southern) 
cluster. In the following, we will refer to these two maser clusters with the name ``J1'',
and will denote the northern and southern ``J1'' clusters with ``J1-N'' and ``J1-S'', respectively.
Most of the water features not belonging to ``J1'' present a NE--SW oriented, 
S-shape distribution (inside the white dashed polygon of Fig.~\ref{mas_dis_1}),
which consists of a NE--SW extended ($\approx$2\arcsec in size, PA$\approx$70\degr) strip of features,
and two smaller ($\approx$0\farcs5 in size), SE--NW oriented lines of masers at the ends
of the NE--SW strip. In the following, we refer to this S-shape pattern of maser
features as ``J2'', and use the names ``J2-SW'' and ``J2-NE'' to indicate the water maser clusters at the
southwestern and northeastern ends of the ``J2'' distribution, respectively. 
We note that while the \Vlsr\ of the ``J2'' masers are relatively
close (within $\pm$10~\kms) to \Vsys, ``J1'' includes
the most blue-~and~red-shifted water features (up to 30~\kms away from \Vsys).

\begin{figure*}
\centering
\includegraphics[width=14cm]{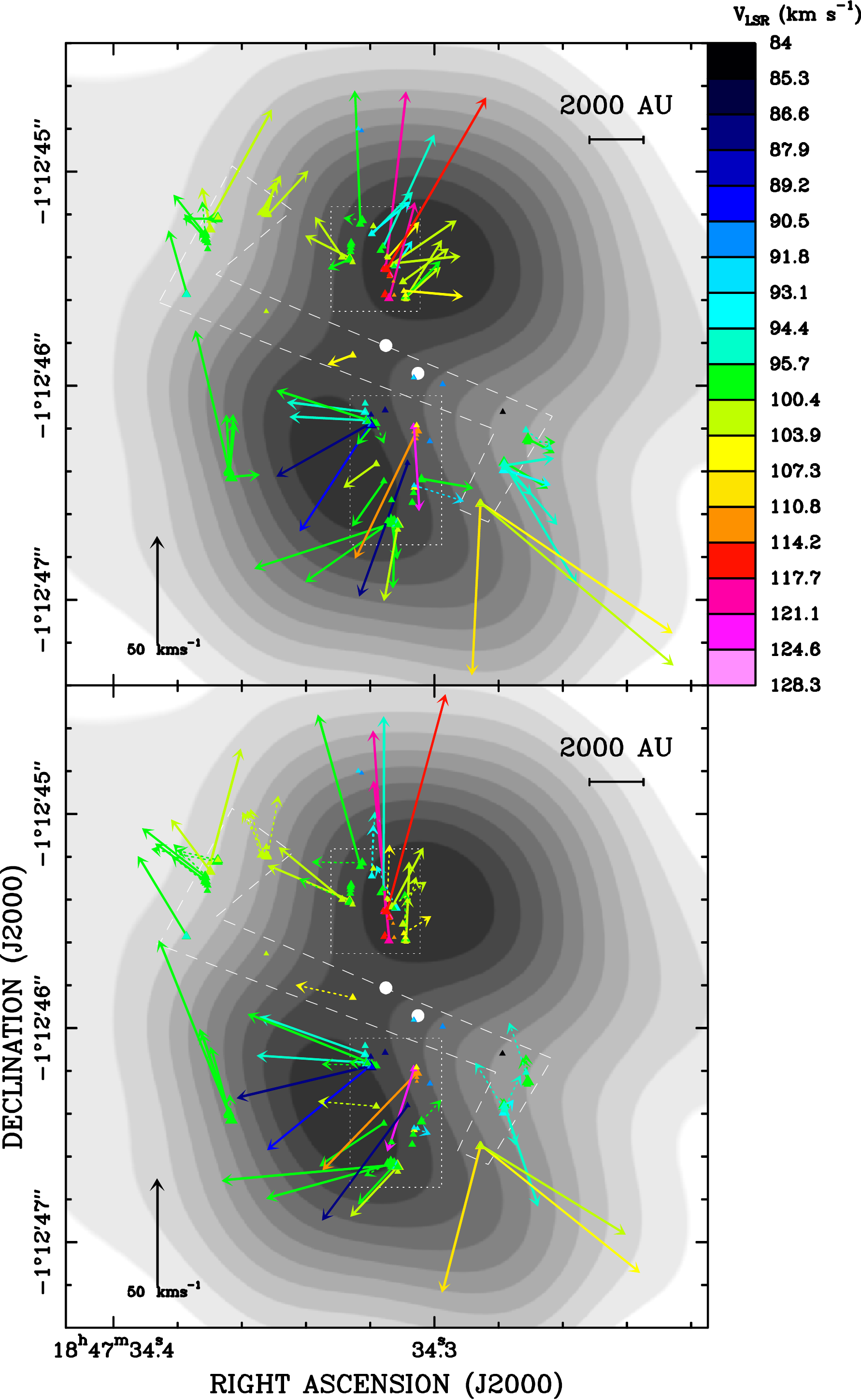}
\caption{\small {\it Upper} and {\it lower} panels show the relative 
(referred to the ``center of motion'') and (methanol-corrected)
absolute proper motions of the 22~GHz water masers, respectively. 
{\it Colored vectors} indicate the direction and the amplitude
of the proper motions. The {\it black vector} in the bottom left corner 
of the panels indicates the amplitude scale of proper motions in kilometer per second.
{\it Dotted vectors} are used for the most uncertain proper motions.
{\it Colors}, {\it symbols}, the {\it grey-scale map}, the {\it white big dots},
the {\it dotted rectangles}, and the {\it S-shape
dashed polygon}, have the same meaning
as in Fig.~\ref{mas_dis_1}.}
\label{wat_pm}
\end{figure*}

We calculated the geometric
center (hereafter ``center of motion'', identified with label \#0 in
Table~\ref{h2o_tab}) of the features with a stable spatial and
spectral structure, persisting over the four observing epochs, and
refer our measurements of proper motions to this point. 
The upper panel of Fig.~\ref{wat_pm} presents the proper motions of the water masers
relative to their ``center of motion''.
The magnitude of the relative proper motions (measured for 70~features and listed in Table~\ref{h2o_tab})
ranges from \ $\approx$10~\kms\  to \ $\approx$120~\kms, with a mean value of
\ 34.0~\kms, and a mean error of \ 7.0~\kms. 
The proper motions of the water masers belonging to the ``J1-N'' and ``J1-S'' clusters 
indicate expansion, since most of the proper motion vectors diverge from the HMC center and are oriented approximately
perpendicular to the pattern outlined by the water masers. 
%In each of the two clusters, we also note that maser features at the cluster border present
%smaller velocities (both along the line-of-sight and on the sky) than features placed 
%more internally inside the cluster. 
Concerning the velocity distribution of the ``J2'' water masers, the few measured proper motions for the features
found in the NE--SW strip suggest that the masers are receding
from the HMC center moving along directions at close angle
with the strip orientation. The water features in the ``J2-SW'' cluster
appear to move consistently towards SW, parallel to the strip orientation.
%, with masers
%more detached from the strip line showing increasingly larger velocities.
In contrast, the proper motions of the ``J2-NE'' maser features present a more scattered 
distribution, with the direction of motion rotating from NE to NW with the maser position
varying from SE to NW.  
%closer to the strip line, to being oriented close to North for the features at the largest separation 
%from the strip. 
Finally, we note that the water maser features not belonging to ``J1'' 
and ``J2'' groups, clustered to the SE of the whole maser distribution, present
motions to the North.

To check the reliability of the relative, water maser proper
motions we follow the same procedure applied in previous studies.
%\citet{Li12}.
%, which
%uses the relatively small, absolute proper motions of the
%6.7~GHz methanol masers (derived in Sect.~\ref{res_met}) to correct the large absolute
%proper motions of the water masers.
%As described in Sect..~\ref{res_met}, we have also derived the absolute proper
%motions of the 6.7~GHz methanol maser features. 
%To date, VLBI observations of 6.7~GHz \citep{san10a,san10b} and 12~GHz
%\citep{mos03,mos10,mos11} methanol masers toward several sources
%indicate that \textbf{these masers usually show small proper motions}, with
%typical velocities of less than 10--15~km~s$^{-1}$. 
Assuming that in \Gd\ (as verified in several sources \citep{Mos10,Mos11,San10a,San10b}) 
methanol masers move (with respect to the star) significantly slower than the water masers, we
can use the absolute proper motion of the methanol masers to correct
the absolute proper motion of the water masers.
The result should be a good approximation of the water maser velocities
with respect to the reference system comoving with the star.
% and derive the water
%maser velocities as approximately seen in a reference system
%comoving with the star. 
The lower panel of Fig.~\ref{wat_pm} shows
the absolute water maser proper motions after subtracting the
average absolute, methanol maser proper motion (evaluated in
Sect..~\ref{res_met}). Note that, within the errors,
the methanol-corrected absolute proper motions of the water masers 
are consistent with the relative proper motions calculated 
with respect to the ``center of motion'' (upper panel of Fig.~\ref{wat_pm}). 
This result supports our assumption
that the ``center of motion'' can be a suitable reference
system to calculate water maser velocities. In the following
discussion, we use the relative velocities of the water masers to
describe the gas kinematics close to the MYSO.

The methanol-corrected absolute proper motions reported 
in the lower panel of Fig.~\ref{wat_pm} should not be confused
with the absolute proper motions of the water maser features, 
derived from the measured absolute proper motion (corrected for the apparent motion)
of feature~\#1 (see Table~\ref{tbl:mas-abs-pos}). 
%The latter has been calculated adding the
%contribution of the parallax, the solar motion with respect to the
%LSR \citep{sch10}, and the differential Galactic rotation, from a
%flat rotation curve with Galactic constants (R$_0 = 8.3\pm0.23$~kpc,
%$\Theta_0 = 239\pm7 $~\kms) as recently determined by \citet{bru11}.
The absolute proper motions differ from the relative proper
motions presented in the upper panel of Fig.~\ref{wat_pm} by a vector of
amplitude \ $\approx$60~\kms\ pointing to NE, which seems
to be the dominant component of all proper motions and reflect the large
systematic deviation affecting the measurement of the absolute proper motions.
%We think that
%this difference could be caused by the large systematic error due to
%the uncertainties in the source distance and solar and Galactic
%motion, and/or by a peculiar velocity of the maser source with
%respect to its LSR reference system. Being severely affected
%by this systematic error, the intrinsic absolute proper motions 
%of water masers cannot be used to describe the gas kinematics
%around the MYSO.
% and we do not show their plot.

%\addtocounter{table}{1}

\subsection{1.6~GHz Hydroxyl Masers}
\label{res_oh}

\begin{figure*}
\centering
\includegraphics[width=17cm]{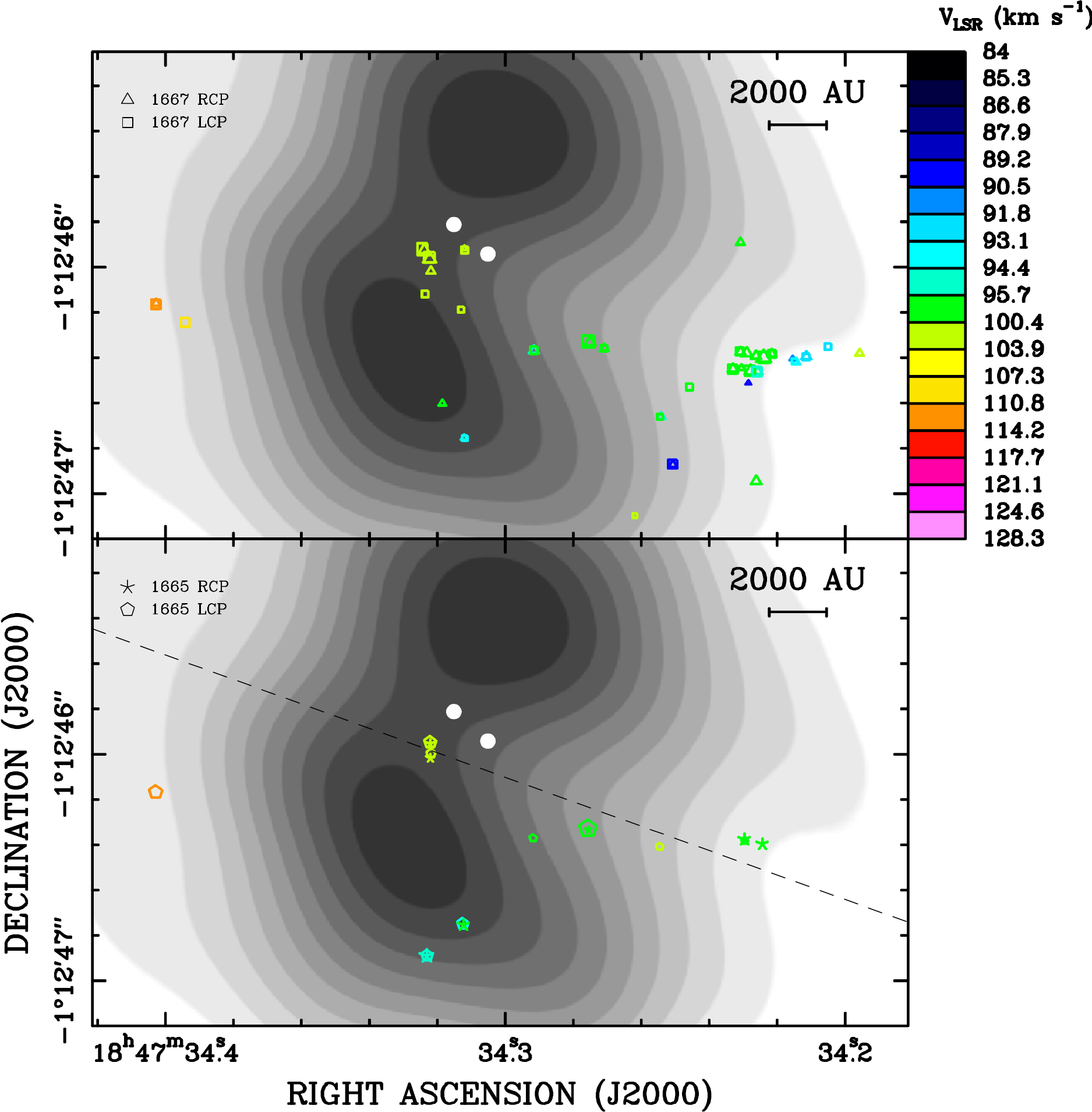}
\caption{\small {\it Lower} and {\it upper} panels show the absolute position 
and LSR velocity of the 1665~MHz and 1667~MHz OH masers, respectively.
For the 1665~MHz masers, the LCP and RCP features are indicated
with {\it empty pentagons} and {\it stars}, respectively.
{\it Empty squares} and {\it empty triangles} indicate the LCP and
RCP features of the 1667~MHz masers, respectively.
Symbol area is proportional to the logarithm of the feature intensity.
 The {\it black dashed line} in the lower panel shows the axis of projection of the OH maser positions
to produce the position-velocity plot of Fig.~\ref{oh_grad}.
{\it Colors}, the {\it grey-scale map}, and the {\it white big dots}, have the same meaning
as in Fig.~\ref{mas_dis_1}.}
\label{mas_dis_2}
\end{figure*}

The lower and upper panels of Fig.~\ref{mas_dis_2} show absolute positions and LSR velocities
of the detected 1665~MHz and 1667~MHz OH maser features, respectively, whose parameters 
are listed in Tables~\ref{oh1665_tab}~and~\ref{oh1667_tab}. The number
of detected features is significantly higher at 1667~MHz (52) than at 
1665~MHz (17), consistent with the fact that the average OH maser intensity is
higher at 1667~MHz (0.9~Jy~beam$^{-1}$) than at 1665~MHz (0.7~Jy~beam$^{-1}$).
For both OH maser transitions, the detections are similarly distributed
between RCP and LCP components, which also show comparable intensities.
The overall OH maser \Vlsr\ interval extends from 89~\kms\ to 111~\kms.
The 1665~MHz and 1667~MHz OH maser spatial distributions are similar,
both extending more along the East-West ($3\arcsec$--$3\farcs5$)
than along the North-South ($\approx$1\arcsec) direction. In particular,
we note that the spatial distribution of both the 1665~MHz and 1667~MHz OH masers 
is not symmetrical with respect to the two VLA continuum sources found at the center of the HMC. 
Both maser transitions populate preferentially the region to the West of the VLA sources,
spreading more or less uniformly up to a maximum offset of $\approx$2\arcsec.
Instead, to the East of the VLA sources, the observed OH maser features concentrate at relatively 
small ($\le$0\farcs5) offsets and very few features are found at large ($\approx$1\farcs5) separation from the HMC center.    
The OH masers present also a regular change in \Vlsr\ with the position, 
with \Vlsr\ increasing from West to East. We discuss further the properties 
of the OH maser spatial and velocity field in \  Sect.~\ref{HMC_exp}.

\begin{table*}[h]
\caption{\label{oh1665_tab} 1665~MHz OH Maser Parameters.}
\centering
\begin{tabular}{rrcrrr}
\hline\hline
\multicolumn{1}{c}{Feature} & \multicolumn{1}{c}{Pol.} & \multicolumn{1}{c}{I$_{\rm peak}$} & \multicolumn{1}{c}{$V_{\rm LSR}$} & \multicolumn{1}{c}{$\Delta~x$} & \multicolumn{1}{c}{$\Delta~y$}  \\
\multicolumn{1}{c}{Number}  &      & \multicolumn{1}{c}{(Jy beam$^{-1}$)} & \multicolumn{1}{c}{(km s$^{-1}$)} & \multicolumn{1}{c}{(mas)} & \multicolumn{1}{c}{(mas)}  \\
\hline
%!comp  Pol. int_av(Jy|beam)  vel_av(km|s)   RA(mas) +/$-$ e_RA(mas)     DEC(mas) +/$-$ e_DEC(mas)  
    1 & LCP &      4.85 &    98.0 &   0.00$\pm$0.00  &   0.00$\pm$0.00  \\
    2 & LCP &      0.86 &   111.4 &  1906.92$\pm$0.35 &   160.94$\pm$0.47  \\
    3 & LCP &      0.57 &   103.7 &   695.99$\pm$0.70 &   380.96$\pm$1.05  \\
    4 & LCP &      0.43 &    93.7 &   552.01$\pm$0.77 &  $-$420.45$\pm$0.94  \\
    5 & LCP &      0.41 &    94.8 &   709.72$\pm$0.70 &  $-$560.67$\pm$0.92  \\
    6 & LCP &      0.23 &   102.9 &   692.67$\pm$0.99 &   330.48$\pm$1.41  \\
    7 & LCP &      0.16 &    96.9 &   241.37$\pm$1.34 &   $-$42.28$\pm$1.94  \\
    8 & LCP &      0.15 &   102.2 &  $-$317.18$\pm$1.50 &   $-$80.13$\pm$1.85  \\
    9 & LCP &      0.14 &    97.9 &  $-$692.10$\pm$1.84 &   $-$53.60$\pm$2.05  \\
   10\tablefootmark{a} & LCP &      0.11 &   109.3 & $-$1757.31$\pm$2.26 &  $-$842.74$\pm$2.71  \\
\hline   
%!comp  Pol.  int_av(Jy|beam)  vel_av(km|s)   RCPA(mas) +/$-$ e_RCPA(mas)     DEC(mas) +/$-$ e_DEC(mas)  
    1 & RCP &      1.49 &    94.8 &   711.06$\pm$0.26 &  $-$560.97$\pm$0.33  \\
    2 & RCP &      0.82 &    97.7 &  $-$690.91$\pm$0.47 &   $-$46.86$\pm$0.56  \\
    3 & RCP &      0.76 &    96.0 &   549.14$\pm$0.40 &  $-$424.47$\pm$0.49  \\
    4 & RCP &      0.73 &    97.6 &  $-$769.73$\pm$0.52 &   $-$67.21$\pm$0.55  \\
    5 & RCP &      0.41 &   103.2 &   696.07$\pm$0.82 &   374.85$\pm$1.12  \\
    6 & RCP &      0.32 &   102.8 &   695.33$\pm$1.13 &   312.30$\pm$1.24  \\
    7 & RCP &      0.24 &    97.6 &    $-$2.33$\pm$1.81 &    $-$4.39$\pm$2.11  \\
\hline
\end{tabular}
\tablefoot{Column~1 gives the feature label number; column~2
indicates the circular polarization of the maser emission;
columns~3~and~4 report the intensity of the strongest spot and 
the intensity-weighted LSR velocity, respectively; 
columns~5~and~6 list the position offsets (with the
associated errors) along the R.A. and Dec. axes, relative to the
feature~\#1 of LCP.\\
\tablefoottext{a}{This single feature is significantly offset to West with respect to the other OH maser detections, and it is not plotted in Figure~\ref{mas_dis_2}.}
}
\end{table*}

\begin{table*}[h]
\caption{\label{oh1667_tab} 1667~MHz OH Maser Parameters.}
\centering
\begin{tabular}{rrcrrr}
\hline\hline
\multicolumn{1}{c}{Feature} & \multicolumn{1}{c}{Pol.} & \multicolumn{1}{c}{I$_{\rm peak}$} & \multicolumn{1}{c}{$V_{\rm LSR}$} & \multicolumn{1}{c}{$\Delta~x$} & \multicolumn{1}{c}{$\Delta~y$}  \\
\multicolumn{1}{c}{Number}  &      & \multicolumn{1}{c}{(Jy beam$^{-1}$)} & \multicolumn{1}{c}{(km s$^{-1}$)} & \multicolumn{1}{c}{(mas)} & \multicolumn{1}{c}{(mas)}  \\
\hline
%comp Pol. int_av(Jy|beam)  vel_av(km|s)   RA(mas) +/$-$ e_RA(mas)     DEC(mas) +/$-$ e_DEC(mas)  
    1 & LCP &      6.20 &    97.8 &   0.00$\pm$0.00  &   0.00$\pm$0.00  \\
    2 & LCP &      3.42 &    97.9 &  $-$777.01$\pm$0.19 &   $-$69.90$\pm$0.19  \\
    3 & LCP &      2.30 &   103.6 &   702.06$\pm$0.21 &   371.33$\pm$0.23  \\
    4 & LCP &      1.65 &    96.2 &  $-$713.49$\pm$0.29 &  $-$128.56$\pm$0.25  \\
    5 & LCP &      1.61 &   103.8 &   731.84$\pm$0.25 &   411.39$\pm$0.29  \\
    6 & LCP &      1.08 &    95.3 &  $-$743.87$\pm$0.38 &  $-$132.20$\pm$0.32  \\
    7 & LCP &      0.99 &    96.9 &  $-$637.47$\pm$0.38 &  $-$122.11$\pm$0.40  \\
    8 & LCP &      0.94 &   111.2 &  1907.56$\pm$0.46 &   164.47$\pm$0.49  \\
    9 & LCP &      0.72 &   108.7 &  1778.70$\pm$0.62 &    84.73$\pm$0.57  \\
   10 & LCP &      0.71 &    89.5 &  $-$372.31$\pm$0.48 &  $-$540.98$\pm$0.49  \\
   11 & LCP &      0.53 &    97.3 &  $-$669.36$\pm$0.56 &   $-$43.81$\pm$0.62  \\
   12 & LCP &      0.41 &    97.0 &  $-$810.71$\pm$0.77 &   $-$54.05$\pm$0.71  \\
   13 & LCP &      0.37 &    96.6 &   239.25$\pm$0.85 &   $-$39.08$\pm$1.04  \\
   14 & LCP &      0.27 &    92.1 &  $-$961.22$\pm$1.22 &   $-$68.13$\pm$1.30  \\
   15 & LCP &      0.26 &   103.1 &   545.39$\pm$1.12 &   404.28$\pm$1.18  \\
   16 & LCP &      0.25 &   100.3 &  $-$446.38$\pm$1.16 &  $-$200.60$\pm$1.26  \\
   17 & LCP &      0.24 &   101.9 &   720.38$\pm$1.42 &   209.70$\pm$1.43  \\
   18 & LCP &      0.23 &    99.0 &   $-$70.31$\pm$1.10 &   $-$29.90$\pm$1.16  \\
   19 & LCP &      0.22 &    92.9 & $-$1057.53$\pm$1.71 &   $-$22.39$\pm$1.63  \\
   20 & LCP &      0.17 &   102.0 &   561.21$\pm$1.73 &   141.02$\pm$1.91  \\
   21 & LCP &      0.17 &    96.0 &  $-$316.07$\pm$1.63 &  $-$332.44$\pm$1.97  \\
   22 & LCP &      0.15 &    94.1 &   545.39$\pm$1.42 &  $-$426.44$\pm$1.68  \\
   23 & LCP &      0.15 &    93.2 &  $-$909.66$\pm$1.86 &   $-$87.36$\pm$1.59  \\
   24 & LCP &      0.12 &   100.6 &  $-$206.02$\pm$1.84 &  $-$768.08$\pm$2.36  \\
\hline
%comp Pol. int_av(Jy|beam)  vel_av(km|s)   RA(mas) +/$-$ e_RA(mas)     DEC(mas) +/$-$ e_DEC(mas)  
    1 & RCP &      6.13 &    97.6 &  $-$773.11$\pm$0.17 &   $-$69.95$\pm$0.16  \\
    2 & RCP &      2.92 &   103.5 &   723.50$\pm$0.28 &   399.02$\pm$0.31  \\
    3 & RCP &      2.25 &   103.2 &   700.38$\pm$0.20 &   366.62$\pm$0.22  \\
    4 & RCP &      1.28 &    96.3 &  $-$715.28$\pm$0.42 &  $-$129.39$\pm$0.35  \\
    5 & RCP &      1.09 &    97.6 &  $-$698.57$\pm$0.62 &   $-$50.59$\pm$0.54  \\
    6 & RCP &      0.91 &    97.3 &    $-$6.42$\pm$0.53 &    $-$2.40$\pm$0.55  \\
    7 & RCP &      0.68 &    97.6 &  $-$740.27$\pm$1.11 &  $-$615.34$\pm$0.87  \\
    8 & RCP &      0.63 &    96.8 &  $-$637.96$\pm$0.61 &  $-$119.64$\pm$0.60  \\
    9 & RCP &      0.61 &    97.8 &  $-$740.02$\pm$0.78 &   $-$62.81$\pm$0.82  \\
   10 & RCP &      0.55 &    92.0 &  $-$961.68$\pm$0.72 &   $-$65.10$\pm$0.66  \\
   11 & RCP &      0.54 &    97.1 &  $-$671.03$\pm$0.59 &   $-$45.12$\pm$0.60  \\
   12 & RCP &      0.46 &    95.4 &  $-$747.35$\pm$0.82 &  $-$131.52$\pm$0.73  \\
   13 & RCP &      0.44 &    95.7 &   240.08$\pm$0.69 &   $-$38.21$\pm$0.84  \\
   14 & RCP &      0.41 &    96.8 &  $-$809.27$\pm$0.94 &   $-$52.88$\pm$0.89  \\
   15 & RCP &      0.38 &   102.8 &   694.81$\pm$0.94 &   311.94$\pm$0.97  \\
   16 & RCP &      0.37 &    98.9 &   $-$70.41$\pm$0.91 &   $-$29.06$\pm$0.94  \\
   17 & RCP &      0.35 &   111.1 &  1907.42$\pm$0.75 &   169.23$\pm$0.92  \\
   18 & RCP &      0.34 &    89.9 &  $-$372.86$\pm$0.72 &  $-$542.73$\pm$0.75  \\
   19 & RCP &      0.34 &    96.6 &  $-$671.38$\pm$1.39 &   438.76$\pm$1.16  \\
   20 & RCP &      0.32 &   102.4 & $-$1196.87$\pm$1.14 &   $-$50.74$\pm$1.33  \\
   21 & RCP &      0.31 &    95.3 &  $-$321.55$\pm$0.91 &  $-$331.49$\pm$1.04  \\
   22 & RCP &      0.27 &    93.3 &  $-$917.20$\pm$1.19 &   $-$89.33$\pm$1.20  \\
   23 & RCP &      0.23 &    96.6 &  $-$675.39$\pm$1.54 &  $-$116.40$\pm$1.32  \\
   24 & RCP &      0.23 &    98.9 &   644.40$\pm$1.10 &  $-$272.34$\pm$1.34  \\
   25 & RCP &      0.17 &    95.4 &   547.94$\pm$1.78 &  $-$424.08$\pm$2.35  \\
   26 & RCP &      0.15 &   102.9 &   546.51$\pm$2.07 &   412.25$\pm$2.21  \\
   27 & RCP &      0.15 &    91.1 &  $-$901.27$\pm$1.60 &   $-$75.20$\pm$1.66  \\
   28 & RCP &      0.13 &    89.8 &  $-$706.07$\pm$1.90 &  $-$182.16$\pm$2.37  \\
\hline
\end{tabular}
\tablefoot{Column~1 gives the feature label number; column~2
indicates the circular polarization of the maser emission;
columns~3~and~4 report the intensity of the strongest spot and 
the intensity-weighted LSR velocity, respectively; 
columns~5~and~6 list the position offsets (with the
associated errors) along the R.A. and Dec. axes, relative to the
feature~\#1 of LCP.
}
\end{table*}

\subsection{6.7~GHz Methanol Masers}
\label{res_met}

The upper panel of Fig.~\ref{mas_dis_1} shows absolute positions and LSR velocities
of the detected methanol maser features, whose parameters are listed in Table~\ref{ch3oh_tab}.
Over four epochs, we detected 85 distinct maser features,
whose intensity ranges from the detection threshold of \ $\approx$40~mJy~beam$^{-1}$
\ up to $\approx$10~Jy~beam$^{-1}$, and with \Vlsr\ varying from 90~\kms\ to 107~\kms.
6.7~GHz masers distribute at distances \ $\la$1\arcsec\ from the two compact VLA continuum sources (at the HMC center),
mainly clustering to the SE, SW and NW of the continuum positions,  
with no 6.7~GHz features detected to the East of the VLA sources.
Comparing the methanol and water maser spatial distributions, we find two clusters
of 6.7~GHz features (red-~and~blue-shifted and placed inside the northern 
and the southern dotted rectangles of Fig.~\ref{mas_dis_1}, respectively)
which have positions and \Vlsr\ similar to 22~GHz features belonging 
to the North-South oriented, ``J1'' group.
A resemblance in positions and \Vlsr\ between the two maser species is
also observed to the SW of the VLA continuum sources, where methanol masers are spread
nearby the ``J2-SW'' water maser group.
We find, however, also a small cluster of 6.7~GHz features to the NW of 
the continuum sources (inside a dotted ellipse in Fig.~\ref{mas_dis_1}),
detached from the area of water maser detection. In the following, we will refer to this 
6.7~GHz maser cluster with the name ``F''.

\begin{figure*}
\centering
\includegraphics[width=13.7cm]{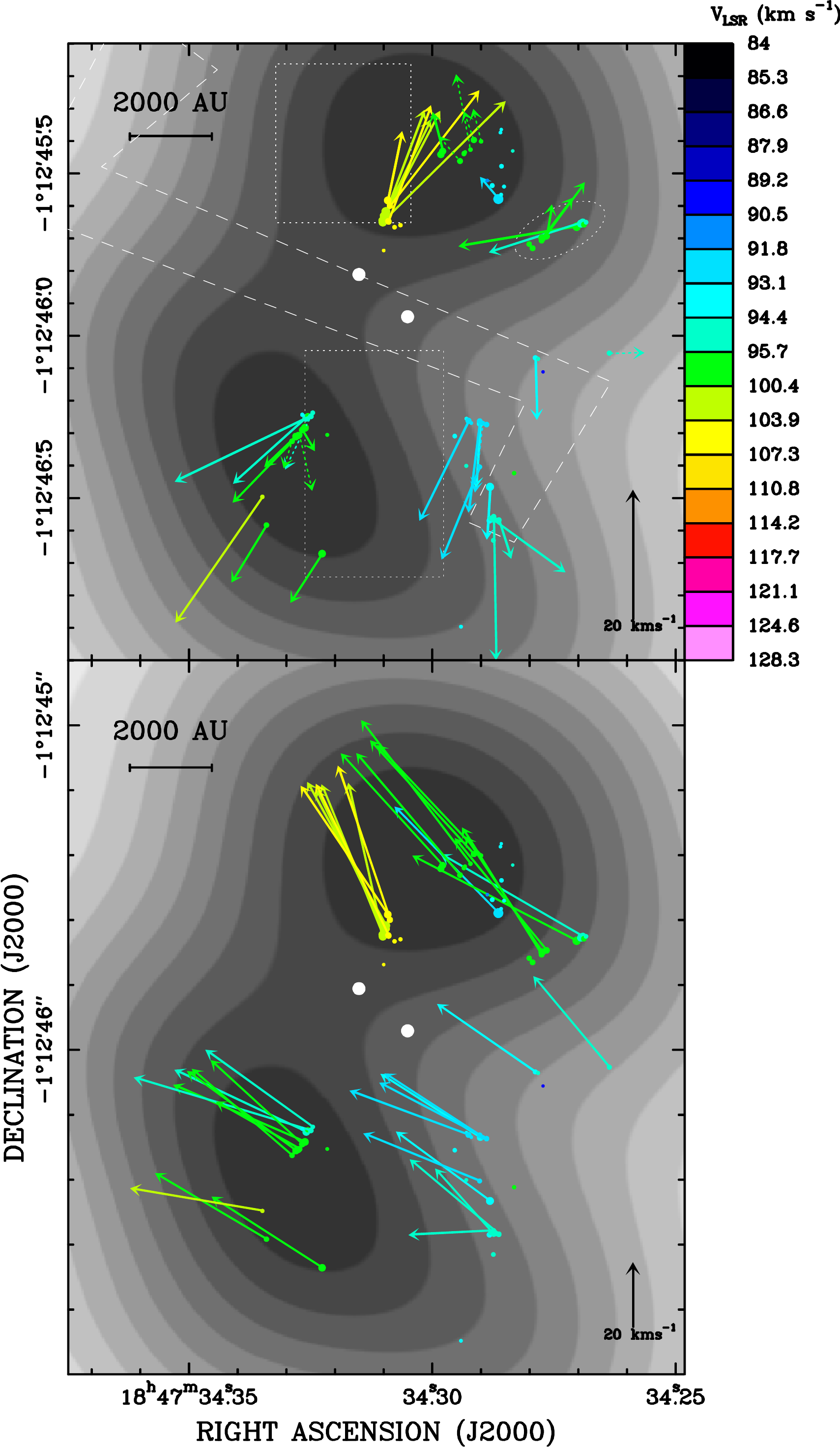}
\caption{\small {\it Upper} and {\it lower} panels show the relative 
(referred to the ``center of motion'') and 
absolute (corrected for the apparent motion, see Sect.~\ref{mas_obs})
proper motions of the 6.7~GHz methanol masers, respectively. 
{\it Colored vectors} indicate the direction and the amplitude
of the proper motions. The {\it black vector} in the bottom right corner 
of the panels indicates the amplitude scale of proper motions in kilometer per second.
{\it Dotted vectors} are used for the most uncertain proper motions.
{\it Colors}, {\it symbols}, the {\it grey-scale map}, the {\it white big dots},
the {\it dotted rectangles}, the {\it S-shape
dashed polygon}, and the {\it dotted ellipse}, have the same meaning
as in Fig.~\ref{mas_dis_1}.}
\label{met_pm}
\end{figure*}

We calculated the geometric
center (hereafter ``center of motion'', identified with label \#0 in
Table~\ref{ch3oh_tab}) of the features with a stable spatial and
spectral structure, persisting over the 4 observing epochs, and
refer our measurements of proper motions to this point. 
The upper panel of Fig.~\ref{met_pm} presents the proper motions of the methanol masers
relative to their ``center of motion''.
The magnitude of the relative proper motions (measured for 40~features and listed in Table~\ref{ch3oh_tab})
ranges from \ $\approx$5~\kms\  to \ $\approx$30~\kms, with a mean value of
\ 13.0~\kms, and a mean error of \ 4.0~\kms. It is interesting to compare
the proper motions of the water and methanol masers. 
The two clusters of 6.7~GHz features in proximity of the ``J1'' H$_2$O maser structure 
(inside the dotted rectangles of Fig.~\ref{met_pm}, upper panel)
have proper motions similarly oriented
as the nearby water masers, with \meth\ features in correspondence of 
the northern (southern) ``J1'' lobe, moving to NW (SE) (see Fig~\ref{jet1_elli}, right panel). 
The ``J1'' water masers, however,
move on average faster than the adjacent methanol masers.
Proper motions of only $\approx$5~\kms\ (pointing to N-NE) are measured for the group
of 6.7~GHz masers offset \ 0\farcs1--0\farcs5 \ to the West of the the ``J1-N'' water maser cluster.  
The methanol masers
with positions intermediate between the ``J1-S'' and the ``J2-SW'' water maser clusters
move preferentially to South, while the motion of the closeby, ``J2'' water masers 
is mainly oriented to SW. Finally, we note that the features in the 6.7~GHz maser cluster ``F''
(inside a dotted ellipse in Fig.~\ref{met_pm}, upper panel) present proper motions in opposite
directions.

The lower panel of Fig.~\ref{met_pm} shows the absolute
proper motions of the 6.7~GHz \meth\ masers. These differ from the relative proper
motions (presented in the upper panel of Fig.~\ref{met_pm}) by a vector, 
corresponding to the average absolute proper motion of the 6.7~GHz masers, with
components of \ $29\pm4$~\kms\ along the right-ascension axis and 
\ $29\pm7$~\kms\ along the declination. As it is clear from Fig.~\ref{met_pm},
the dispersion ($\approx$10~\kms\ in each Equatorial coordinate) 
% non-weighted standard deviation: 8.5 km/s along the RA, and 11.8 km/s along the DEC
of the 6.7~GHz maser, 
absolute proper motions from the mean is small, suggesting, as noted in Sect.~\ref{res_wat}, 
that a large systematic effect dominates our measurement of absolute proper motions.
Since for the 6.7~GHz \meth\ masers, this systematic deviation  (with amplitude 
of \ $\approx$41~\kms) overwhelms the average amplitude ($13$~\kms) of
the relative proper motions,  its value can be adequately estimated as
the average of the 6.7~GHz maser, absolute proper motions.
As already explained in Sect.~\ref{res_wat}, the absolute proper motions
of the 22~GHz \wat\ masers presented in the lower panel of Fig.\ref{wat_pm} have been corrected
for this systematic effect.
 
%\addtocounter{table}{2}

\section{A Double-Jet System inside the \Gd\ HMC}
\label{jets_exp}

\subsection{The ``J1'' Water Maser Group}
\label{discu_jet1}

\begin{figure*}
\centering
\includegraphics[width=16cm]{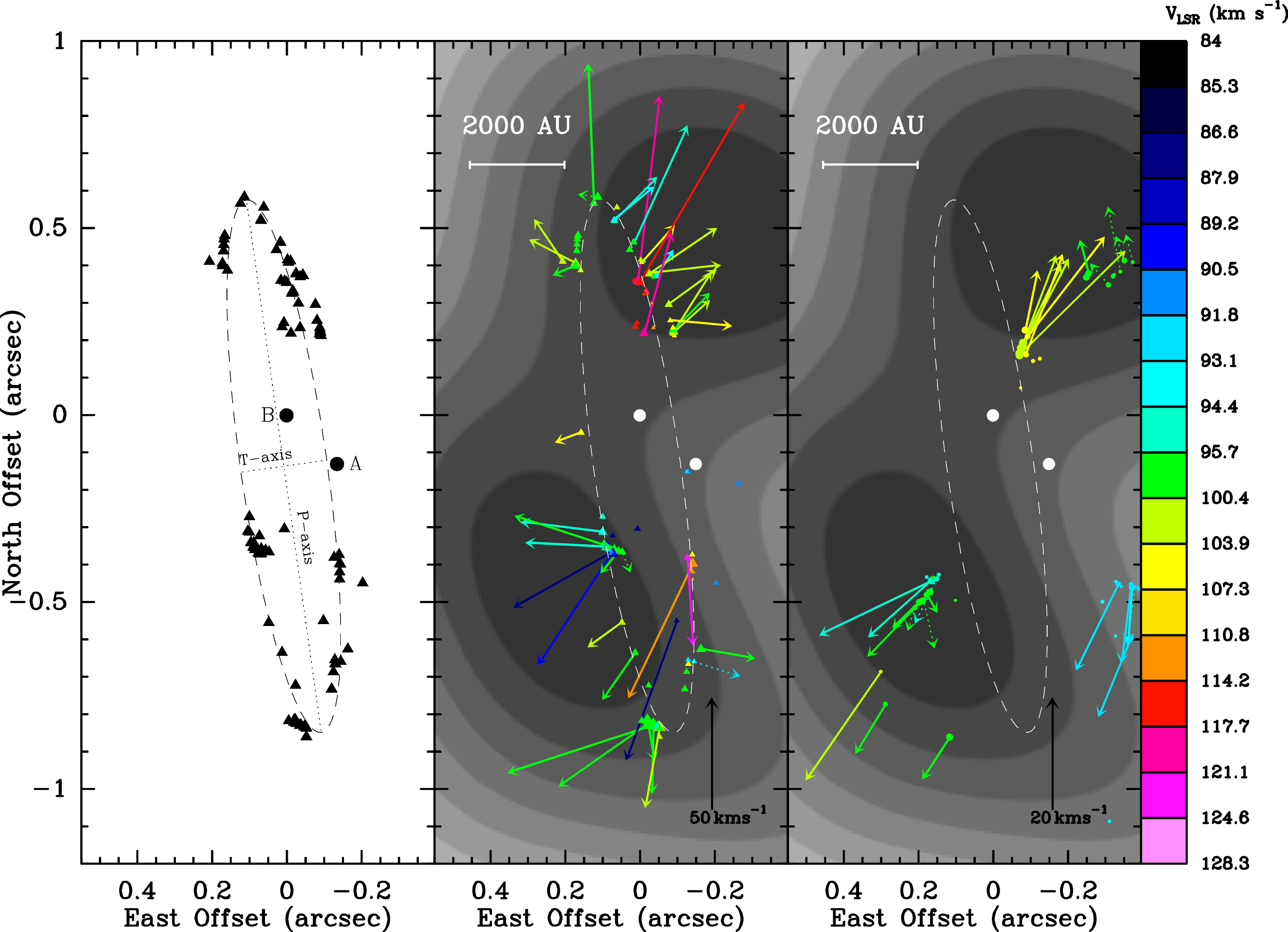}
\caption{\small Each of the three panels shows the same area of the sky around the ``J1'' water maser group. 
Maser positions are always refered to the VLA source ``B'' (see text).
In each panel, the {\it dashed line} shows the best-fit ellipse to the
distribution of the ``J1'' water masers.  \ 
{\bf Left~panel:}~{\it Black triangles} denote the positions of the ``J1'' water masers. 
The {\it dotted lines} trace the major (labeled ``P-axis'') and minor (``T-axis'')
axis of the best-fit ellipse. {\it Black dots} give the positions of the VLA sources ``A'' and ``B''. \
{\bf Central~and~Right~panels:}~{\it Colored triangles} and {\it dots} show the positions of the
water ({\it central~panel}) and methanol ({\it right~panel}) masers, respectively. Symbol area is proportional to the
logarithm of the feature intensity.
{\it Colors}, the {\it grey-scale map}, the {\it white big dots}
have the same meaning as in Fig.~\ref{mas_dis_1}.
{\it Colored vectors} indicate the direction 
and the amplitude of the maser proper motions. The {\it black vector} in the bottom right corner 
of the panels indicates the amplitude scale of proper motions in kilometer per second.
{\it Dotted vectors} are used for the most uncertain proper motions.
}
\label{jet1_elli}
\end{figure*}

In Sect.~\ref{res_wat}, we noted the symmetrical spatial and velocity
distributions of the water masers belonging to the ``J1-N'' and ``J1-S'' clusters.
Figure~\ref{jet1_elli} (left panel) shows that these distributions can be well fit
with an ellipse. 
%For this fit, we have selected the subset of
% ``J1'' water features with \Vlsr\ closer to \Vsys\ (i.e., 90~\kms\ $\le$ \Vlsr\ $\le$ 110~\kms),
%excluding the most blue-~and~red-shifted features mainly found offset from the
%border
%\emph{to the interior} of the ``J1'' maser distribution (see Fig.~\ref{mas_dis_1}, lower panel). 
The fit parameters are: \ 1)~the position on the sky of the ellipse center; 
\ 2)~the major and minor axes; \ 3)~the PA (East from North) of the major axis. 
The best-fit position of the ellipse center 
is determined to be \ 0\farcs14$\pm$0\farcs03
\ South (PA = 177\degr$\pm$5\degr) of the VLA source ``B'' (i.e. the NE source; see Fig.~\ref{mas_dis_1}), which 
is closer to the ``J1'' center than the VLA source ``A''.
The best-fit major and minor axes are
\ 1\farcs44$\pm$0\farcs1 \ and \ 0\farcs24$\pm$0\farcs02, respectively, with 
PA = 8\degr$\pm$1\degr.

An elliptical spatial distribution is expected if the masers
traced a circular ring inclined with respect to the plane of the sky. However, 
%in such a case, the simplest scenario where 
assuming that the masers also move on the ring plane 
(as, for instance, in an expanding ring),
this scenario
is clearly to be ruled out on the basis of the comparison of the  
fit ellipse eccentricity and the 3-D maser velocities.
 The ratio between the minor and the major 
axis of the best-fit ellipse to the ``J1'' maser distribution
gives an inclination angle of the plane of the putative maser ring
with respect to the plane of the sky of 
\ $i_{\rm cir}$ = $\arccos$(0\farcs24/1\farcs44) $\simeq 80$\degr,
indicating that the maser circle should be seen close to edge-on.
On the other hand, if masers move on the ring plane, 
the ring inclination angle can be also evaluated by the
ratio of maser velocity components projected along the line-of-sight, $V_z$, and 
along the ellipse minor axis, $V_T$.
% maser velocity
%components is to be constant and equal to the ring inclination
%angle. 
Averaging on all the (39) ``J1'' features with measured proper motions, since
in general \ $V_T \gg V_z$, the ring inclination angle
as determined from the maser velocities is \ 2\degr$\pm$2\degr, 
implying that the maser motion should
be seen face-on. That contrasts with the elongated
maser spatial distribution, and we have to
rule out the ring interpretation (with masers moving on the ring plane).

%Figure~\ref{jet1_vel} suggests a
%correlation between the maser velocity components and the distances  
%parallel to the ``J1'' axis, even better defined for the
%fastest maser features found closer to this axis. 
If the H$_2$O  masers in ``J1'' are accelerated by a bipolar jet
directed approximately North-South, that could explain
the elongated elliptical distribution of the masers and
the pattern of proper motions mainly diverging from the ellipse center.  
%the velocity-position correlation can be explained
%assuming the ``J1'' axis is the direction on the sky along which a
%(proto)stellar jet is accelerating the water maser motion.
%Since the global ``J1'' maser spatial distribution is well fit
%with an ellipse, 
The putative location of the MYSO
driving this jet likely coincides with the
ellipse center. As previously noted, the best-fit position of the ellipse center 
lies only \ $\approx$0\farcs14 to the South of source ``B'' (see Fig.~\ref{jet1_elli}, left panel). 
%Since such an offset is a significant fraction of the best-fit ellipse
%major semi-axis (0\farcs724$\pm$0\farcs03), and it is remarkably directed
%very close to the jet axis, one cannot exclude that it is real.
Moreover, based on the cm-wavelength SED analysis,
\citet{Ces10} reason  that the  source ``B''
%using a power law with spectral index 1.3. Besides, the
%attempt of these authors to fit the
%SED with a model of an homogeneous \ion{H}{II} region was unsuccessful 
%because it is required
%a source size much larger than the observed value.
%These results from the SED fit suggest that 
%the source ``B'' 
could be a thermal jet 
rather than a compact \ion{H}{II} region.
We conclude that source ``B'' could be excited
by the same (proto)stellar jet driving the ``J1'' water maser motions.

One can estimate the momentum rate of the jet
from the average distance of the H$_2$O masers from the MYSO and the
average maser velocity. Assuming a H$_{2}$ pre-shock density \ 
$n_{\rm H_{2}} = 10^{8}$~cm$^{-3}$, as predicted by excitation models of 22~GHz masers \citep{eli89}, the
momentum rate in the water maser jet is given by the expression:
\begin{equation}
\label{MR_WM}
\dot{P} = 1.5 \times 10^{-3} \, V_{10}^{2} \, R_{100}^{2} \, (\Omega/4\pi) \ \,  M_{\odot} \, \mathrm{yr}^{-1} \, \mathrm{km} \, \mathrm{s}^{-1}
\end{equation}
Here \ $V_{10}$ \ is the average maser velocity in units of
10~km~s$^{-1}$, $R_{100}$ the average distance of water masers from
the YSO in units of 100~AU, and $\Omega$ the solid angle of the jet.
This expression has been obtained by multiplying the momentum rate
per unit surface transferred to the ambient gas ($n_{\rm H_{2}} \,
m_{\rm H_{2}} \, V^{2}$), by $\Omega R^{2}$, under the assumption
that the jet is emitted from a source at a distance $R$ from the
masers within a beaming angle $\Omega$. To estimate the average
distance and velocity of the water masers tracing the jet, we
use all the (39) maser features with measured proper motions
belonging to the ``J1'' group. The
resulting average distance is 0\farcs47 (or $\approx$3700~AU) and the average
speed is 36~km~s$^{-1}$. Using these values, the jet momentum rate
is $\dot{P} \simeq 27 \, (\Omega/4\pi)$~$M_{\odot}$~yr$^{-1}$~km~s$^{-1}$.

The ratio between the minor and the major axis of the best-fit ellipse 
 can be used to estimate the jet semi-opening angle \ $\theta_j \simeq 10\degr$,
which, assuming that the jet is symmetric about its axis, 
corresponds to a jet solid angle \ $\Omega \simeq 0.1$~sterad.
Using such a value for \ $\Omega$, the jet momentum rate results 
\ $\dot{P} \simeq 0.2$~$M_{\odot}$~yr$^{-1}$~km~s$^{-1}$.  
%According to the observational results of \citet{Lop10,Beu02b}, 
%such an high value of jet momentum rate would correspond to a
%YSO luminosity \  $>10^4~L_\odot$.

% Most of the ``J1'' water masers are found in correspondence
% of the two (northern and southern) CH3CN emission peaks,
% which could trace dense clumps on which jet inmpinges.
% The orientation of the two CH3CN peaks might mark a flattened structure.
% 
% Thus the ``J1'' jet is still confined within a dense cocoon
% within a radius <= 1" from the MYSO. Dynamical time of 10^3 yr.
%% The ``J1'' jet is possibly still confined inside the dense cocoon of dust and gas,
%% which might explain why its geometry is so well traced by water masers (excited only at
%% high densities). In particular note how many of the ``J1'' water masers are found
%% at position of higher CH3CN emission at \Vlsr = \Vsys, which could trace denser gas 
% 
Figures~\ref{mas_dis_1}~and~\ref{wat_pm} show that, in both the ``J1-N'' and ``J1-S'' clusters,
most of the water maser features are found projected close to the peaks of the bulk
CH$_3$CN (12-11) line emission. A possible interpretation
is that
% the strongest CH$_3$CN emission marks two dense clumps, symmetrically
%displaced to the North and the South of the MYSO, onto which the ``J1'' jet
%impinges, shocking the molecular gas and originating the water masers.
%At the ``J1'' extremities, the continuous arrow-shaped pattern 
%of the water masers suggests that the maser emission is tracing the working
%surface of the jet head, 
the H$_2$O masers are excited at the interface between the jet and the 
surrounding dense molecular core.
The profile outlined by the water masers in the ``J1-N'' and ``J1-S'' clusters
reminds that of the bow-shocks associated with Herbig-Haro objects \citep{Har11}, produced 
by protostellar jets when they impinge against the dense circumstellar material.
This scenario suggests that the  ``J1'' jet is still 
confined inside a dense cocoon of dust and gas within a distance \ $\le$1\arcsec\
from the MYSO, 
 which is further indicated by the fact that, despite our sensitive observations,
we have failed to reveal a molecular outflow
along the ``J1'' (North-South) direction up to angular scales of several \ 10\arcsec \citep{Ces11}.
Using the average value of \ $\approx$20~\kms\ for the jet-parallel
component of maser velocities, and the jet major semi-axis of \ 0\farcs7, 
one derives a short dynamical time scale of \ $1.3 \, 10^3$~yr.

\subsection{The ``J2'' Water Maser Group}
\label{discu_jet2}
% Elongated from NE to SW, with features at the SW end, moving to SW.
% The water maser strip tracing the jet at P.A. =
% The VLA source "A", crossed by the water maser strip, might also be shock-ionized free-free emission 
% A second jet, possibly emerging from VLA~A, with the NE lobe perturbated
% by another flow traveling from South to North.
%%
Looking at the ``J2'' water maser distribution (see Figs.~\ref{mas_dis_1}~and~\ref{wat_pm}),
the NE--SW elongated strip of water masers and the motion of the water features in the ``J2-SW'' cluster, parallel 
to the strip orientation, suggests the presence of another collimated outflow, at PA$\approx$70\degr.
We speculate that the water masers in the strip mark an ionized jet, amplifying the 
free-free continuum on the background: in this view the emission from source ``A'', 
which lies along the maser strip (see Fig.~\ref{mas_dis_1}),
could be tracing the jet close to the MYSO. 
However, the proposed scenario cannot readily account for the direction of motion of water masers
in the ``J2-NE'' group, where several maser features move to the North rather than parallel to
the direction (PA$\approx$70\degr) of the putative jet. Note also that to the South (offset by $\approx$1\arcsec)
of the ``J2-NE'' cluster, another cluster of water maser features (inside the dashed circle in Fig.~\ref{envi_grad}, lower panel) with proper motions directed close to North is found.
The motion of these features do not seem to be consistent with the jet interpretation for the ``J2'' masers.
Section~\ref{discu_envi} propones a qualitative explanation of the more scattered velocity distribution
for the masers to the East of the VLA source ``A'', considering the dynamical interaction of the HMC gas with its environment.
In the following  we discuss the scenario that the ``J2'' masers trace a jet.

Repeating the calculation performed for the ``J1'' maser group in Sect.~\ref{discu_jet1}, 
one can estimate the momentum rate of the ``J2'' outflow from the average maser distance and velocity.
Using the peak of the SMA 1.3~mm continuum (see Fig.~\ref{envi_grad}, lower panel), which,
within the measurement error, is coincident with the VLA source ``A'', as the best-guess position of the MYSO, 
and considering only the ``J2-SW'' water masers, tracing a collimated flow,
the average distance and speed are \ 0\farcs62 and \ 38~\kms, respectively. 
Taking the water masers of the ``J2-NE'' group,
with more scattered direction of motion,  the average distance and speed are \ 1\farcs1 and \ 19~\kms.
In deriving these values, we have averaged over a similar number of persistent maser features, 12 \ and \ 13 \ for the southwestern
and northeastern cluster, respectively. For the ``J2-NE'' maser group, the separation from the MYSO is about twice as great as that
of the masers in the ``J2-SW'' cluster, and the average speed is approximately half as much.
Then, using Eq.~(\ref{MR_WM}) for the jet momentum rate, the values determined using the southwestern
and northeastern maser clusters are in good agreement. That lends support to our argument that the {\em same} jet
is driving the motion of both ``J2'' maser lobes. The momentum rate
of the ``J2'' jet is \  $\dot{P} \approx 50 \, (\Omega/4\pi)$~$M_{\odot}$~yr$^{-1}$~km~s$^{-1}$,
where \ $\Omega$ is the solid angle of the jet.

In an attempt to search for outflows emerging from the HMC in \Gd, 
%%(traced by the CH$_3$CN(12-11)~K=4 line (see Figure~\ref{mas_obs})),
 \citet{Ces11} have used the SMA to map the emission of the
CO(2-1) line with an angular resolution of
\ $\approx$0\farcs8.
% and a velocity resolution of \ 1.0~\kms. 
%The $^{12}$CO(2-1) emission pattern around the \Gd\ HMC
%is quite complex for (blue- and red-shifted) LSR velocities of modest
%separation, 5-10~\kms, from \Vsys. In this \Vlsr\ range,
%one can identify four main $^{12}$CO(2-1) emission features, respectively 
%offset to the East, to the West, to NE and SW of the HMC.
%However, at larger velocities (10~\kms $ \le | $\Vlsr $ - $ \Vsys $ | \le $20~\kms)
%the relative intensity of the $^{12}$CO(2-1) features offset to NE and SW
% with respect to the East and West features, 
%decrease, and 
The high-velocity (10~\kms $ \le | $\Vlsr $ - $ \Vsys $ | \le $20~\kms)
$^{12}$CO(2-1) emission presents a bipolar structure,
with red-~and~blue-shifted emission peaks offset to the East and 
the West of the HMC, respectively \citep[see][Fig.11]{Ces11}.
%with only a residual, weak emission spur to SE. 
The line crossing the two $^{12}$CO(2-1) emission peaks
%offset to the East (red-shifted) and to the West (blue-shifted) 
%of the \Gd\ HMC 
forms a small angle ($\le$20\degr)
with the elongation axis (at PA$\approx$70\degr) of the ``J2'' water maser distribution.
The separation on the sky of the two emission lobes, $\approx$5\arcsec,
is larger than the extension, $\approx$2\arcsec, of the ``J2''
maser group along its major axis.  That suggests that the water masers in ``J2''
trace the root of a collimated outflow which has propagated throughout the HMC
accelerating the core gas up to 20~\kms\ (along the line-of-sight) at separations of $\sim$10$^4$~AU from the
powering MYSO. 
That the observed bipolar (East-West) distribution of the high-velocity $^{12}$CO(2-1) emission could mark a collimated outflow has been also discussed by
\citet{Ces11}. Using the CH$_3^{13}$CN(12-11) lines, which show a well defined \Vlsr\ gradient (at PA = 68\degr) consistent in orientation  and amplitude with that seen in  $^{12}$CO(2-1) and, in addiction, being optically thin, permit a reliable estimate of the whole flow mass, 
these authors derive a value for the outflow momentum rate
of \ $\dot{P}$ = 0.3~$M_{\odot}$~yr$^{-1}$~km~s$^{-1}$. From the expression quoted
above for the momentum rate of the jet driving the ``J2'' water masers, such a jet would require only a
small solid angle  \ $\Omega\approx$0.07~sterad (corresponding to a semi-opening angle \ $\theta_j\approx$9\degr),
to be powerful enough to sustain the molecular flow observed at larger scales.

The momentum rate derived for both the ``J1'' and ``J2'' water maser jets
is of the order of a few tenths of \ $M_{\odot}$~yr$^{-1}$~km~s$^{-1}$.
Single-dish studies with 10\arcsec\ angular resolution of molecular outflows
in massive star-forming regions indicate that such values of momentum rates are
close to the upper limits measured for outflows from MYSOs
\citep[][and refs. therein]{Lop09}.
% and typical of source bolometric luminosities $\ge$10$^6$~$L_{\odot}$, 
%significantly higher than the value of \ 3$\times$10$^5$~$L_{\odot}$ \citep{Ces94} estimated for \Gd.
On the other hand, interferometric observations at sub-arcsecond angular resolution
have identified a small sample of MYSOs (HH~80-81 \citep{Mar98}; IRAS~16547$-$4247 \citep{Rod08};  IRAS~20126$+$4104 \citep{Mos11}) which, although being of moderate
 luminosity (bolometric luminosity in the
range \ 10$^4$--10$^5$~$L_{\odot}$), eject very powerful, collimated thermal jets and molecular outflows (momentum rates of the order of  0.1~$M_{\odot}$~yr$^{-1}$~km~s$^{-1}$). These results suggest
that the most compact and youngest (dynamical time \ $\le$10$^4$~yr) outflows
in massive star-forming regions (observable only with sub-arcsecond angular resolution)
are also among the most powerful and collimated ones observed. 
Note that the dynamical time scale derived for the  ``J1'' jet using the water
masers is only of \ $1.3 \, 10^3$~yr, and that of the ``J2'' jet deduced from the
 (spatial and velocity) separation of the $^{12}$CO(2-1) lobes is  \ $4 \, 10^3$~yr \citep{Ces11}. 
% High-angular resolution observations of a larger sample of outflows
%from MYSOs are required to ascertain if the youngest flows are intrinsically more powerful and collimated. If that were %confirmed, the outflow evolution would
%proceed similarly for low-mass and high-mass (proto)stars. 

%( We caution that the expression used to calculate the momentum rate 
%%depends on the square of both the maser velocity and the maser separation from the star,
%and thus it varies with the fourth power of the source distance. As an example, an excess of the
%adopted source distance (7.9~kpc) by 30--50\% would make the estimate of the momentum rates
%wrong by a factor of 3--5. 

\subsection{Methanol and OH Maser Dynamics: Jet-driven Expansion of the HMC}
\label{HMC_exp}

%6.7~GHz methanol masers close to the ``J1'' water masers are also
% tracing gas entrained in the jet. This is one of the best cases
% for 6.7~GHz masers tracing outflow walls.
As noted in Sect.~\ref{res_met}, from the comparison of the positions and 
proper motions of the 6.7~GHz methanol (Fig.~\ref{met_pm})
and 22~GHz water (Fig.~\ref{wat_pm}) masers, one sees that
the two clusters of 6.7~GHz masers near the ``J1'' jet (see Fig.~\ref{jet1_elli}, central and right panels) show
consistent velocities with the adjacent water masers, despite the lower speeds.
The positions and velocities of the 6.7~GHz masers near the ``J1'' jet suggest 
that this maser emission also traces gas entrained in the jet. To explain the lower
velocities, one could argue that the 6.7~GHz masers would emerge from gas
at larger separation from the jet axis than the water masers. 
Concerning the overall distribution of the 6.7~GHz masers,
Figure~\ref{met_pm} (upper panel) evidences that most of them are found within a few tenths of arcsec
(or a few 1000~AU) from the ``J1'' water masers (i.e. inside the two dotted rectangles of Fig.~\ref{met_pm});
in particular, 6.7~GHz masers are also detected at intermediate positions between the ``J1-S'' and
the ``J2-SW'' clusters (see Fig.~\ref{mas_dis_1})
%, between the southern dotted rectangle and the southwestern
%extremity of the S-shape dashed polygon). 
The  proximity to the jets may indicate
 that the 6.7~GHz maser excitation could be favoured
by the outflow activity. One can speculate that the mid-infrared photons, required for pumping the 6.7~GHz masers \citep{Cra02},
could more easily penetrate into the dense HMC gas in proximity of the outflows, taking advantage of the cavities
induced by the outflowing gas.
The proposed association of the 6.7~GHz masers with outflows, based 
on the similarity of the 3-D velocities of the methanol and water masers, agrees
with the results 
%of a survey of sites of linearly distributed 6.7~GHz methanol masers,
%in typical outflow diagnostics as the shocked H$_2$ 2.12~\mum\ and SiO lines 
by \citet{DeB03} and \citet{DeB09}, which find that
in most cases the methanol masers 
are elongated parallel to jets traced by the H$_2$ and SiO emissions, suggesting
a physical association of the 6.7~GHz masers with the corresponding outflows.

%In Sect.~\ref{discu_jet1} we have discussed the origin and the motion of the two %clusters of 6.7~GHz methanol
%masers (placed inside the northern and the southern dotted rectangle of Fig.~\ref{mas_dis_1}) found nearby the ``J1'' water masers, suggesting that we are seeing HMC gas relatively close
%to the ``J1'' jet, pushed into the jet direction. 
Looking at the overall distribution of 6.7~GHz maser velocities,
Fig.~\ref{met_pm} (upper panel) shows that
most of the 6.7~GHz maser proper motions are diverging from the HMC center, 
with minor deviations from the radial direction.
%the average direction of motion of 6.7~GHz maser clusters generally forming a small angle ($\le$30\degr) with the vector
We have assumed the HMC center to be the midpoint between the VLA sources ``A'' and ``B''.
%to the maser cluster. 
Altogether, the pattern of 6.7~GHz proper motions marks the expansion (at an average speed of $\approx$10~\kms) of the HMC gas close to the ``J1'' and ``J2'' jets.
% blowing the interior 
%of the core along two, roughly perpendicular directions. 
Observing the \Gd\ HMC with SMA at sub-arcsecond angular resolution, \citet{Gir09} detected an inverse P-Cygni profile
in the spectrum of the \ C$^{34}$S (7-6) line, suggesting infall at velocity of a few \kms\ of the core envelope 
at radii of \ $\approx$10$^4$~AU. Although the 6.7~GHz masers trace comparable radial distances, in the range \ 10$^3$--10$^4$~AU, the pattern of their 3-D velocities presents very limited clues to infall. 
It may be significant that the only inward motions of the 6.7~GHz masers are observed for the two features with the largest distance from the water maser jets (inside the dotted ellipse in Fig.~\ref{met_pm}, upper panel).
If the 6.7~GHz masers are excited mostly in proximity of the ``J1'' and ``J2'' jets, their average, outward motion might not be representative of the motion of the HMC bulk gas, since the gas kinematics at large separation from the jets is not
adequately sampled.

\begin{figure*}
\centering
\includegraphics[width=14cm,angle=0.0]{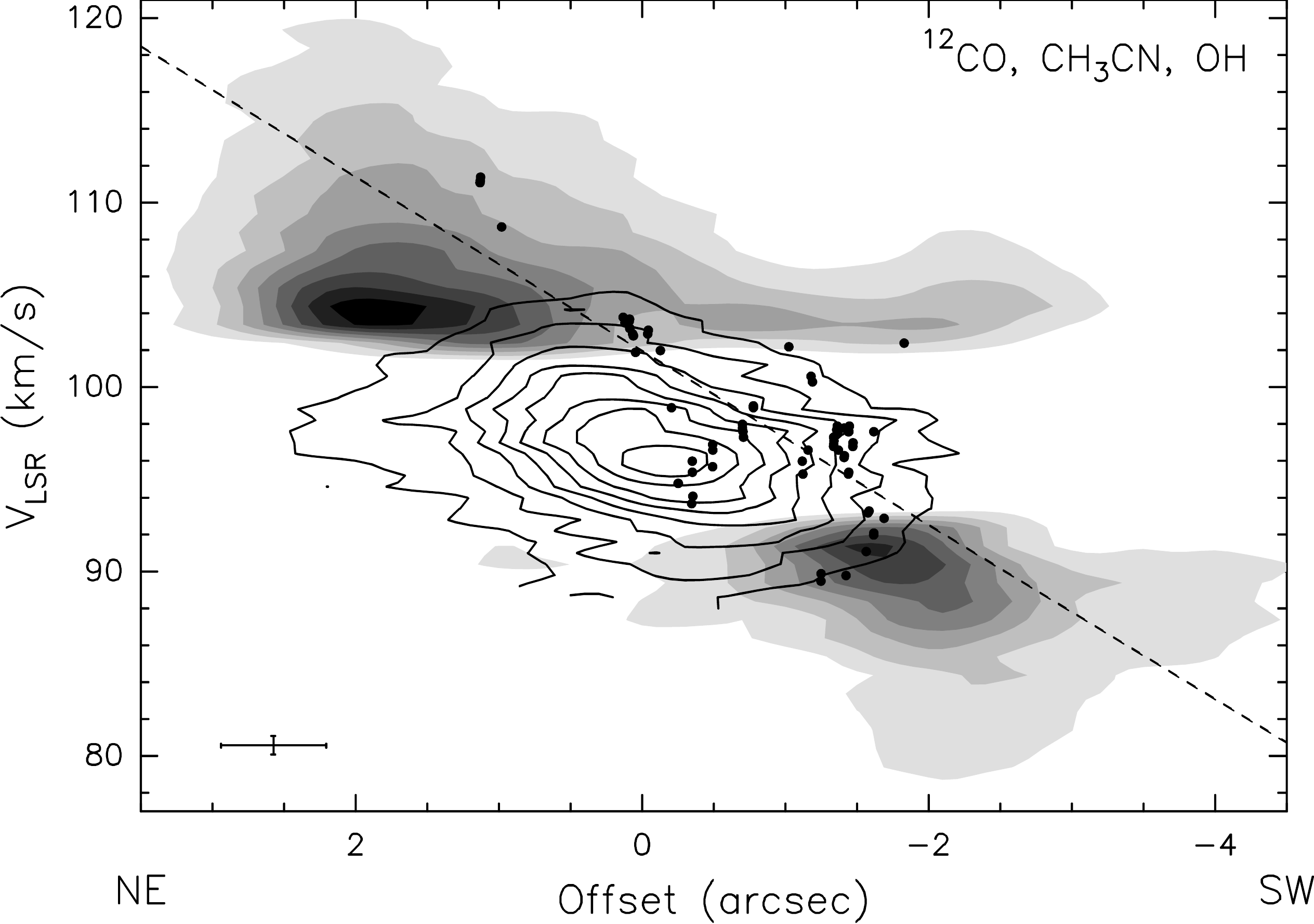}
\caption{\small 
Position-velocity plots of the $^{12}$CO(2--1) (grey scale) and
CH$_3$CN(12--11) K=4 line emission obtained with the SMA by Cesaroni et al.
(2011). The offset is measured along the direction with \ PA = 68\degr\ passing
through the HMC (see Fig.~\ref{mas_dis_2}) and is measured from the phase center of the SMA observations
($\alpha$(J2000)=18$^{\rm h}$47$^{\rm m}$34\fs315,
$\delta$(J2000)=--01\degr12\arcmin45\farcs9), assumed positive toward NE.   
Contour levels increase from \ 0.2 \
in steps of \ 0.2~Jy~beam$^{-1}$ \ for CH$_3$CN and from \ 0.24 \ in steps of \ 0.24~Jy~beam$^{-1}$ \
for $^{12}$CO. The cross in the bottom left denotes the angular and spectral
resolutions of the SMA data.
%Plot of the OH 1.6~GHz maser \Vlsr\ versus the positions projected along the line at PA = 68\degr
%crossing the OH 1667~MHz~and~1665~MHz phase-reference features (which have absolute positions different by less than 2~mas, 
%see Table~\ref{tbl:mas-abs-pos}). The position offset is taken positive towards NE. The two different OH maser transitions
%and polarizations are plotted using different symbols, as indicated in the upper right corner of the panel.
The solid dots represent the 1665~MHz and 1667~MHz OH maser features detected in our VLBA observations.
The {\it black dashed} line gives the best linear fit of the OH maser \Vlsr\ versus corresponding positions.
.
} 
\label{oh_grad}
\end{figure*}

Figure~\ref{mas_dis_2} illustrates that the OH 1.6~GHz masers have an East-West elongated distribution
with an enough regular increase of maser \Vlsr\ with position varying from West to East. The \Vlsr\ gradient 
traced by the OH masers has a similar orientation as the CH$_3$CN \Vlsr\ gradient across the
\Gd\ HMC revealed by \citet{Ces11}. Figure~\ref{oh_grad} presents the OH maser \Vlsr\ versus the
corresponding positions projected along the direction of the CH$_3$CN gradient (PA = 68\degr). 
The maser velocities are well correlated
with the positions (correlation coefficient = 0.8) with a gradient of \ 4.7$\pm$1.6~km~s$^{-1}$~arcsec$^{-1}$.
Inspecting the position-velocity plot of the $^{12}$CO(2-1) line, observed with the SMA by 
\citet{Ces11}, one notes that a similar (both in orientation and in amplitude) 
\Vlsr\ gradient is traced by the high-velocity wings of the $^{12}$CO(2-1) emission (offset in projected position
by \ 4\arcsec--5\arcsec \ in correspondence of the maximum \Vlsr\ separation of \ 20--30~\kms, see Fig.~\ref{oh_grad}). Since, as discussed in Sect.~\ref{discu_jet2}, the $^{12}$CO(2-1) line is likely excited in the collimated, East-West oriented, molecular outflow driven by the ``J2'' water maser jet, we postulate that the observed OH 1.6~GHz masers originate from the same outflow. The similarity of the \Vlsr\ gradient detected
in the high-velocity $^{12}$CO(2-1) emission wings and in the OH masers can indicate
that the OH masers are excited relatively close to the outflow axis (with respect
to the transverse size of the outflow), where the HMC gas is pushed by the jet
to relatively higher velocities. That could explain why in Fig.~\ref{oh_grad} the OH maser \Vlsr\ are offset 
from those of the CH$_3$CN, which should trace the velocity of the HMC bulk gas.

It is interesting to note that, whereas the ``J1'' jet appears to be responsible for the excitation and the motion of most of the 6.7~GHz masers (as discussed above), the OH 1.6~GHz masers seem to be mainly associated with the ``J2'' jet.
%Pumping models of the
%two maser transitions predict similar physical excitation conditions \citep{Cra02}, so that the main difference in the %environments of the ``J1'' and ``J2'' jets causing the preferential production of one of the two masers is likely to be %of chemical nature, i.e. the insufficient abundance of the unseen masing molecule. 
%Actually, some models of outflow chemistry predict that the abundance of CH$_3$OH
%decreases with time, since high-temperature gas-phase reactions transform it into
%simpler molecule as OH. Since the ``J2'' jet is more evolved than the ``J1'' one,
%that could account for the lower (lager) abundance of CH$_3$OH (OH).
These two maser types are both
radiatively excited and models predict similar physical excitation conditions \citep{Cra02},
except for relatively high gas temperature ($\approx$150~K), and densities
($10^{7}$~cm$^{-3}$ $ \le n_{\rm H_{2}} \le 10^8$~cm$^{-3}$), for which OH 1.6~GHz masers are predicted
to thermalize. 
%while CH$_3$OH 6.7~GHz masers should keep elevate brightnesses. 
%the most favourable range of values of gas temperature and (number) density, $n_{\rm H_{2}}$, \
%for strong maser action being \ $\la$100~K and \
%$10^{5}$~cm$^{-3}$ $ \le n_{\rm H_{2}} \le 10^8$~cm$^{-3}$, respectively.
From CH$_3$CN observations, \citet{Bel05} derive a temperature and a density for the HMC gas
of \ $\ga$100~K and \ $\sim$10$^7$~cm$^{-3}$, which is close to the predictions
of the models for OH 1.6~GHz maser quenching. In Sect.~\ref{discu_jet1}, on the basis of the size
and the shape of the water maser distribution, we have postulated that the ``J1'' jet has not yet
emerged from the MYSO natal cocoon, which could be too dense and warm for efficient pumping of the OH 1.6~GHz masers.
On the other hand, the more extended and evolved ``J2'' jet  could have blown the densest portions of the
HMC gas away from its axis, reducing the average gas density to  \ $n_{\rm H_{2}} < 10^7$~cm$^{-3}$, suitable for
exciting the OH 1.6~GHz masers. A reduced gas density, however, cannot explain the relatively 
few 6.7~GHz masers observed towards the ``J2'' jet, since this maser emission should be strong
even at densities \ $\la$10$^7$~cm$^{-3}$. One possibility is that the 
environments of the ``J1'' and ``J2'' jets present chemical differences
affecting the abundance rates between  CH$_3$OH and OH.
%differ chemically too, 
%the abundance of the CH$_3$OH (OH) molecule being higher for the ``J1'' (``J2'') jet.

\section{Discussion}
\label{discu}

\subsection{``J1'' and ``J2'' Jets and the \Vlsr\ Gradient across the HMC}
\label{discu_grad}

% The orientation of the ``J2'' jet is very close to the Vlsr gradient
% detected in CH3CN and other high-density tracers. The momentum rate
%of CH3(13)CN is comparable with that evaluated using the water masers.
%That suggests that the observed Vlsr gradient is due to outflowing gas,
% rather than rotation as sugggested by Cesaroni et al., as also favoured
% Araya et al.. 
One of the most notable findings of previous interferometric (PdBI, SMA) observations
towards \Gd\ is the discovery of a well defined \Vlsr\ gradient 
traced in several transitions of high-density tracers (remarkably CH$_3$CN, see Fig.~\ref{envi_grad})
and also in the $^{12}$CO(2-1) line, a typical outflow tracer.
The interpretation of this velocity gradient, in terms either of a rotating toroid or a collimated outflow, is still under debate. In the following, we discuss the relevance
of the VLBI observations of the water masers to shed light on this issue.

\subsubsection{The Origin of the ``J1'' Jet}

One of the main difficulty to interpret the \Vlsr\ gradient in terms of
(pseudo-)Keplerian rotation is the absence of a jet oriented perpendicular to the
\Vlsr\ gradient, as predicted by star-formation models. We have detected however 
the ``J1'' water maser distribution, which traces a collimated jet traveling
along a direction forming an angle of $\approx$60\degr\ with the gradient orientation.
According to the interpretation by \citet{Ces11}, the \Gd\ HMC is a massive (several 100~$M_{\odot}$),
large rotating toroid feeding a cluster of MYSOs embedded at the center of the core.
One could expect that the accrection disks of these MYSOs are aligned along directions
at close angle with the \Vlsr\ gradient orientation. Then it is plausible that the ``J1'' jet,
forming an angle of $\approx$30\degr\ with the direction perpendicular to the \Vlsr\ gradient,  
is associated with one of these MYSO. 
\citet{Oso09} have recently reproduced the dust spectrum and the ammonia line 
emission of the \Gd\ HMC with a model of an infalling envelope onto a massive star
undergoing an intense accretion phase,
deriving a central star mass of \ 20--25~M$_{\odot}$ and an age of \ 3--4 $ \, 10^4$~yr. 
Note that the modeled protostar age is more than one order of magnitude larger than the 
dynamical time derived for the ``J1'' jet of \ $1.3 \, 10^3$~yr. That would
lead us to rule out the hypothesis that the ``J1'' jet is associated with the
accreting protostar, unless our estimate of the ``J1'' 
dynamical time is not severely underestimated. 

If the ``J1'' jet is precessing, that would explain both the underestimation
of the jet time scale and why the ``J1'' jet is not oriented perpendicular to the \Vlsr\ gradient.
In the literature, it exists at least one, well-studied case, the MYSO IRAS~20126+4104, 
which emits a jet precessing at an angle as large as 45\degr\ \citep{She00}.
The observation of the two VLA compact sources ``A'' and ``B''
at the center of the \Gd\ HMC might
be hinting at a massive binary system, and  
the tidal interaction between the disk and a companion star in a non-coplanar orbit 
is one possible mechanism for jet precession \citep{Ter99,Bat00}.
We do not find any hint of a jet along ``J1'' in the SMA CO maps by \citet{Ces11}.  
The present data towards \Gd\ do not allow to discuss the case for precession
in a more quantitative way. Future ALMA observations
in thermal continuum and line emissions, achieving an angular resolution $\loa 0\farcs1$,
could permit to identify the circumstellar disk around each component of the (putative) binary
system, to verify if the disk planes are misaligned with respect to the binary orbit, and to determine
the binary separation.

\subsubsection{``J2'' Jet Interaction with the HMC}  
 
The ``J2'' water maser distribution (see Figs.~\ref{mas_dis_1}~and~\ref{wat_pm}) 
is elongated along a direction (PA $\approx$70\degr) in good agreement with the orientation (PA = 68\degr) of the \Vlsr\ gradient detected in \Gd\ with the SMA in different lines of CH$_3$CN  \citep{Ces11}. When discussing 
the outflow interpretation for the \Vlsr\ gradient, \citet{Ces11} derive a momentum rate of \ $\dot{P}$ = 0.3~$M_{\odot}$~yr$^{-1}$~km~s$^{-1}$ from the optically thin CH$_3^{13}$CN(12-11) line. 
This value is consistent with our estimate of  \ $\dot{P}$ \
for the ``J2'' water maser jet (see Sect.~\ref{discu_jet2}) for a beaming angle  \ $\Omega\approx$0.07~sterad.
%In Sect.~\ref{discu_jet2}, we have seen that the putative jet responsible for the
%acceleration of the ``J2'' water masers, even if collimated within a small
%solid angle, could easily sustain such a powerful molecular flow. These two facts
We thus favour the interpretation of the  CH$_3$CN \Vlsr\ gradient in terms
of a compact and collimated outflow. 

\citet{Ces11} point out two major difficulties for the outflow interpretation
of the \Vlsr\ gradient.
The first one is associated with the too large flow parameters (in particular
the momentum rate) if compared with the results of single-dish surveys
of massive outflows. As already discussed in Sect.~\ref{discu_jet2}, 
high-angular resolution observations of a few, well studied MYSOs suggest that, 
analogously to what observed in low-mass protostars, 
the youngest and most compact outflows of massive (proto)stars are also intrinsically more powerful.
Thus, the results of single-dish surveys, biased towards more evolved outflows, could 
systematically underestimate the flow parameters at the earliest evolutionary phases (dynamical time \ $\le 10^4$~yr). 
The second concern of \citet{Ces11} is that the
flow dynamical time ($4 \, 10^3$~yr) would be too short with respect to the time 
needed to form typical hot core species -- such as methyl cyanide -- according
to some theoretical models \citep{Cha92}. The formation
route of these models requires that the gas spends at least several 10$^3$~yr at temperatures above 100~K, since the process is powered by the evaporation of 
H$_2$O and other non-refractory species at 100~K, followed by a chain of high-temperature gas-phase reactions to transform simple carbon-bearing molecules into second-generation complex organics. However, these high-temperature models are presently challenged
by the grain-surface models, which predict that the organics
may be formed abundantly on the grain surfaces through mild photochemistry at 
20--40~K \citep{Gar06,Gar08}.The grain-surface pathway takes several 10$^4$~yr at 20--40~K, and such timescales and temperatures are attainable both in the envelope before the material approaches the star and inside the disk (should
it be present). Once formed on the grains, these first-generation organics can be returned to the gas phase by thermal evaporation or liberated by non-thermal processes such as shocks in the outflow walls. The whole formation route requires that the material spends only {\em a few 100~yr} at high-temperatures, an order of magnitude less than the
dynamical time scale of the outflow supposed to be at the origin of the observed
\Vlsr\ gradient in the \Gd\ HMC.

Although the observed water maser kinematics is 
consistent with the outflow scenario,
the maser data themselves are not clear-cut enough to rule out
alternative interpretations. In Sect.~\ref{discu_jet2},
we noted that the water maser velocities to the East of the
VLA sources ``A'' and ``B'' are not collimated along the
axis of the ``J2'' maser distribution, which
might cast some doubts on these masers to belong to a collimated outflow,
and on the efficacy of the ``J2'' flow to produce a well-defined
\Vlsr\ gradient across the HMC. We postpone the discussion of this problem to Sect.~\ref{discu_envi},
where we show how also the anomalous velocities of these features can be explained in the
contest of the jet interpretation for the ``J2'' masers.

\subsection{Interaction of the HMC with its Environment}
\label{discu_envi}

\begin{figure*}
\centering
\includegraphics[width=10.7cm]{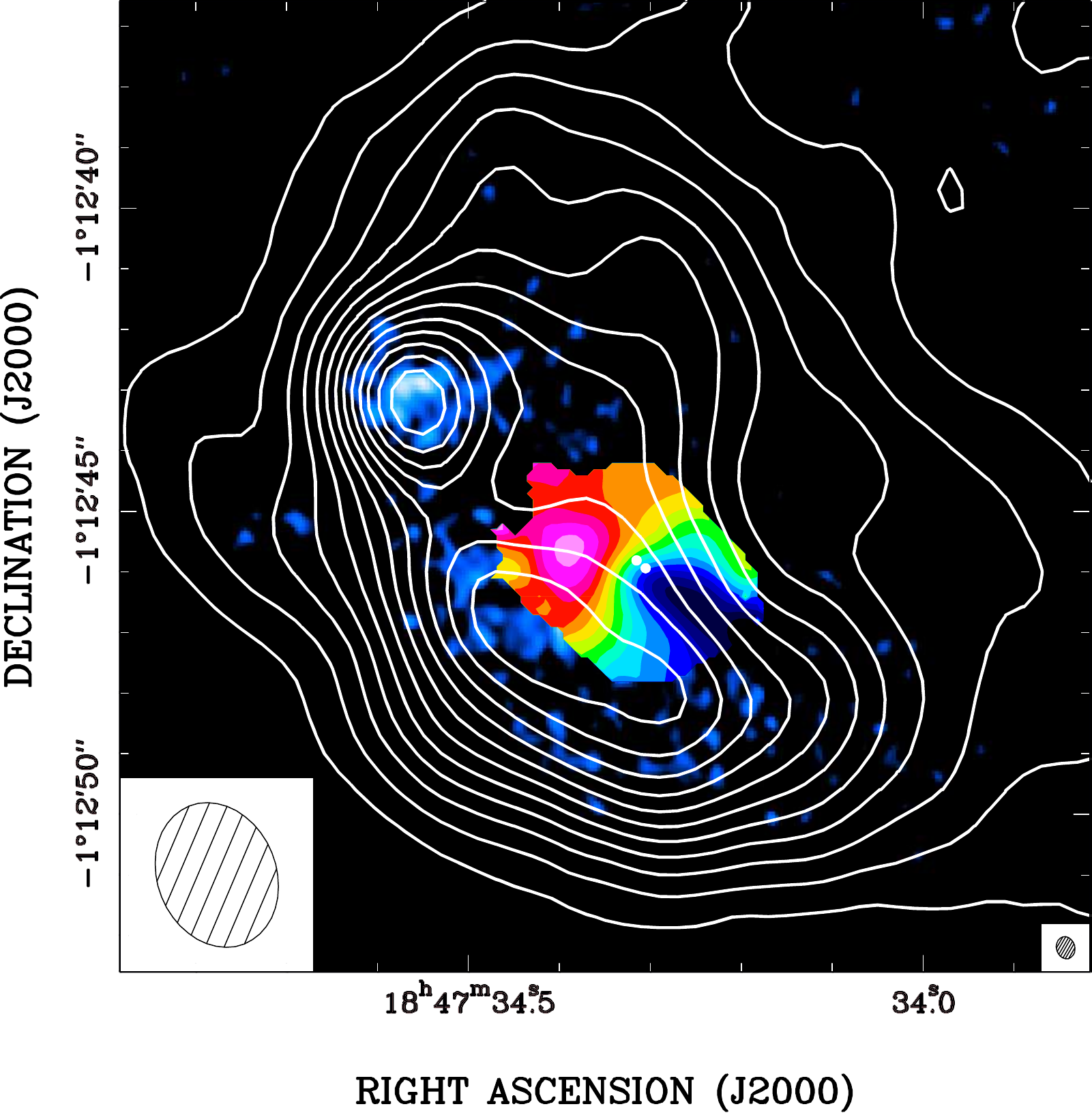}
\includegraphics[width=12.3cm]{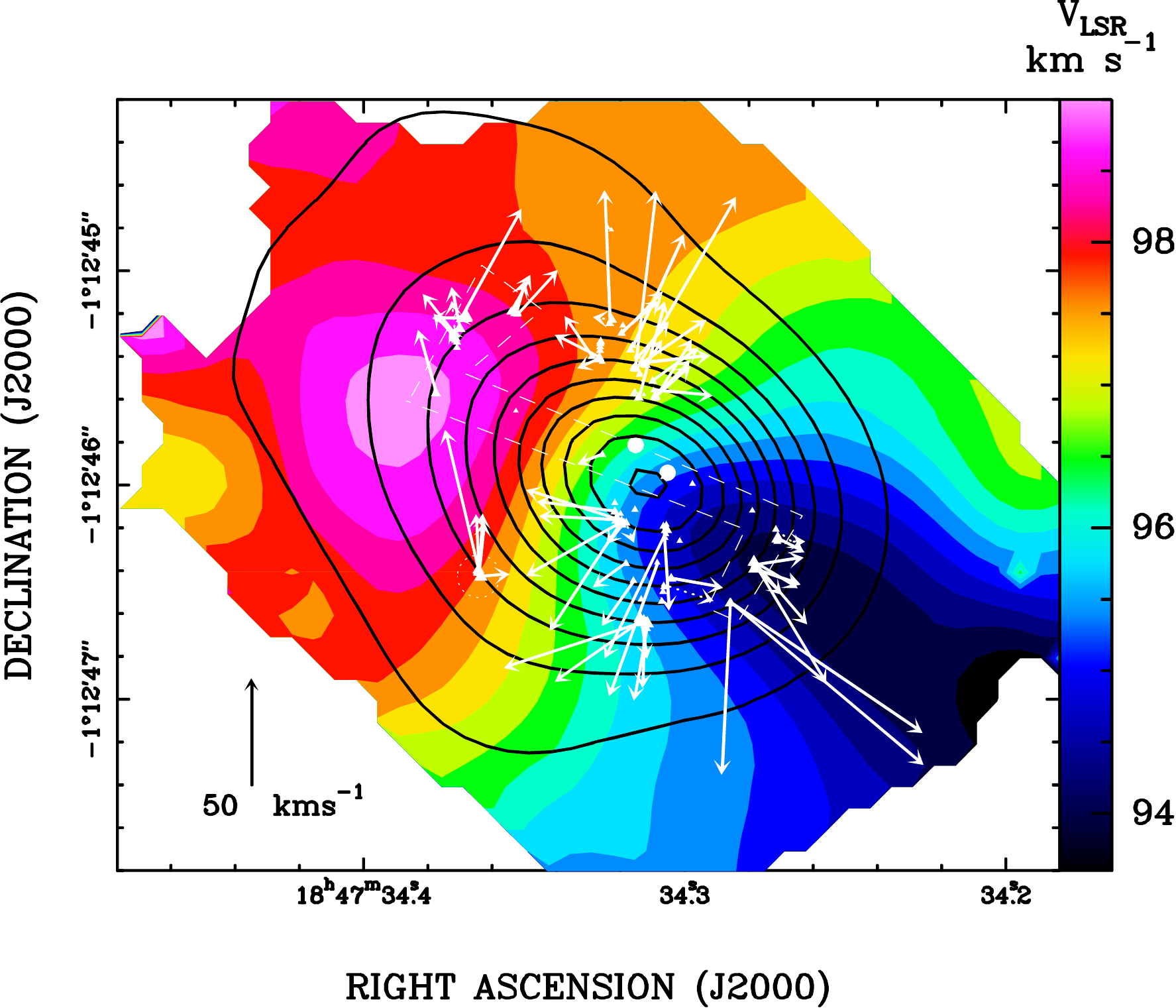}
\caption{\small The \Gd\ HMC ({\it lower panel}) and its environment ({\it upper panel}). 
{\bf Upper~panel:} The {\it blue-tone} image reproduces the VLA B-Array 1.3~cm
continuum observed by \citet{Ces98}, with blue hue varying logarithmically from a minimum level of 2~mJy~beam$^{-1}$
to a maximum of 7~mJy~beam$^{-1}$. The {\it white contour} plot gives the VLA combined C-~\&~D-Array, 1.3~cm
continuum from \citet{Ces94b}: plotted contours range from \ 5~mJy~beam$^{-1}$ to \ 70~mJy~beam$^{-1}$ by steps 
of \ 5~mJy~beam$^{-1}$. The beams of the VLA B-Array and \ the merged C-~\&~D-Array are shown in the inserts in the bottom
right and left corner of the panel, respectively. \  {\bf Lower~panel:} Zoom on the HMC. Overlay of the map of the 1.3~mm continuum
emission ({\it contours}) on that of the CH$_3$CN(12-11) line velocity ({\it color scale}). Contour levels range
from \ 0.1~Jy~beam$^{-1}$ to \ 2~Jy~beam$^{-1}$ by steps of \ 0.2~Jy~beam$^{-1}$. The {\it white dots} denote
the two VLA compact sources detected by \citet{Ces10}. {\it White filled triangles} and {\it white vectors}
give the absolute positions and relative proper motions (with respect to the ``center of motion'', see Sect.~\ref{res_wat}) of the 22~GHz water masers. 
Triangle area is proportional to the logarithm of the maser intensity. The amplitude scale for proper motions is indicated by the {\it black arrow} in the bottom left corner
of the panel. The {\it S-shape dashed polygon} encompasses the ``J2'' water maser distribution and 
the {\it dotted circle} surrounds the isolated cluster of water masers belonging to neither the ``J2'' nor the ''Jet~1''
group.} 
\label{envi_grad}
\end{figure*}

In the upper panel of Fig.~\ref{envi_grad}, we compare the velocity map of the HMC obtained
from the CH$_3$CN molecule with the free-free continuum emission imaged at
1.3~cm by \citet{Ces94,Ces10}. Two continuum maps are shown: one
obtained by merging the C and D configurations of the VLA, the other with the
B configuration. Two peaks of emission are seen in the former image, with the
one to the SE being located very close to the HMC. In the B-array map only
the peak to the NE is still visible: this is very likely tracing the position
of the O6 star ($2\times10^5~L_\odot$; \citet{Ces10}) responsible
for the extended free-free emission seen in Fig.~\ref{envi_grad}, upper panel. The UV photons
escaping from the surroundings of the star could also ionize the surface of
the HMC, thus causing the second, more shallow peak of free-free continuum
observed with the C+D arrays. This scenario strongly suggest that the star
ionizing the HII region and the HMC are close in space and not only in
projection on the plane of the sky.
Further support to this hypothesis is lent by the shape of the millimeter
continuum emission, shown in the bottom panel of Fig.~\ref{envi_grad}. Clearly, the
contour map looks slightly elongated in the NE--SW direction, with
the emission extending towards NE and decreasing more sharply to the SW.
Such an asymmetry can be explained with heating of dust on the
side of the HMC facing the luminous star ionizing the HII region.

We believe that the proximity between the ionizing star and the HMC, besides
explaining the morphology of both the free-free and thermal dust continuum
emission, can also justify the directions of proper motions in the ``J2-NE'' 
H$_2$O maser group. If the ``J2'' masers are tracing a bipolar outflow,
the NE lobe of this can be strongly affected by the interaction with the
stellar photons, possibly enhancing the level of turbulence of the outflowing
molecular gas through heating and photoevaporation \citep{Hen09}.
Albeit impossible to quantify on the basis of the available data, the
effect of these phenomena on the velocity field of the outflowing gas
is likely significant and could explain the anomalous maser proper motions
observed.

It is also interesting to note that a significant enhancement of turbulence
could also explain why no CH$_3$OH maser and only few OH masers
are detected on the NE side of the HMC. These types of masers have measured
diameters $\ga10$~mas, much greater than those of H$_2$O masers ($\sim0.5$mas),
which suggests that also the typical amplification paths are larger.
Therefore, OH and CH$_3$OH masers are also those most affected by an
alteration of the gas velocity field, consistent with the observed lack
of emission.

In conclusion, we believe that scattered proper motions of the ``J2-NE'' cluster
of water masers cannot be used to discard the hypothesis
that such masers are tracing the NE lobe of a bipolar outflow oriented NE--SW.

\section{Conclusions}
\label{conclu}

This work reports on multi-epoch VLBI observations of the H$_2$O 22~GHz and CH$_3$OH 6.7~GHz
masers, and single-epoch VLBA of the OH 1.6~GHz masers towards the HMC in \Gd.
%The gas kinematics of the HMC is well sampled by the three maser species.
%Water maser emission is strong (up to $\approx$100~Jy~beam$^{-1}$) and an elevated number (173) of maser features
%is detected across the HMC. 
Distributed over the HMC, we detect 173 H$_2$O, 85 CH$_3$OH, and 69 OH maser features;
proper motions are measured for 70 H$_2$O and 40 CH$_3$OH, persistent features.
The magnitude of the water maser, relative proper motions 
ranges from \ $\approx$10~\kms\  to \ $\approx$120~\kms, with a mean value of
\ 34.0~\kms, and a mean error of \ 7.0~\kms. 
%Methanol masers are weaker (up to $\approx$10~Jy~beam$^{-1}$) than water masers, but we still
%detect a significant number (85) of features. 
The magnitude of the methanol maser, relative proper motions 
ranges from \ $\approx$5~\kms\  to \ $\approx$30~\kms, with a mean value of
\ 13.0~\kms, and a mean error of \ 4.0~\kms.
%Concerning the OH masers, the number
%of detected features is significantly higher at 1667~MHz (52) than at 
%1665~MHz (17), which is consistent with finding the average OH maser intensity
%higher at 1667~MHz (0.9~Jy~beam$^{-1}$) than at 1665~MHz (0.7~Jy~beam$^{-1}$).

Water masers present a symmetric spatial distribution with respect to the HMC center,
where two nearby (0\farcs2 apart), compact, VLA sources (named ``A'' and ``B'')
are detected by \citet{Ces10}. A first group of water masers, named ``J1'', consists
of two bow-shape distributions, offset to the North and the South of the VLA sources
by $\approx$0\farcs5, with the bow tip pointing to the North (South) for the northern (southern) 
cluster. The ``J1'' distribution is well fit with an elliptical profile (major and minor axis of
\ 1\farcs44$\pm$0\farcs1 \ and \ 0\farcs24$\pm$0\farcs02, respectively, with the major axis
PA (East from North) = 8\degr$\pm$1\degr), and the maser proper motions mainly diverge
from the ellipse center, with average speed of 36~\kms. 
These findings strongly suggest that the ``J1'' water maser group traces
the heads of a young (dynamical time of \ $1.3 \, 10^3$~yr), powerful (momentum rate of \ $\simeq$0.2~$M_{\odot}$~yr$^{-1}$~km~s$^{-1}$), collimated (semi-opening angle \ $\simeq$10\degr)
jet emerging from a MYSO located close (within $\approx$0\farcs15) to the VLA source ``B''. 

Most of the water features not belonging to ``J1'' present a NE--SW oriented, 
S-shape distribution, which consists of an NE--SW extended 
($\approx$2\arcsec in size, PA$\approx$70\degr) strip of features,
and two smaller ($\approx$0\farcs5 in size), SE--NW oriented lines of masers at the ends
of the NE--SW strip. We denote this S-shape, water maser group with ``J2''.
The few measured proper motions for water masers belonging to the NE--SW elongated strip indicate a motion receding
from the two VLA continuum sources (located approximately at the center of the strip), along directions at close angle
with the strip orientation. The water features distributed at the southwestern ``J2'' end
move all to the SW, parallel to the strip orientation.
Instead, at the northeastern end of the strip, the velocities of the water maser features point in different directions. 
%with direction of motion rotating from NE to North for increasing maser distance from the strip line. 
The elongated distribution of the ``J2'' group and the direction of motion, approximately parallel to the direction of elongation, of most ``J2'' water masers suggests that the masers are tracing another collimated outflow, emitted from
a MYSO placed near the VLA source ``A''.
%The average distance (0\farcs62) from the MYSO of the masers at the ``J2'' southwestern end is approximately
%the half of that (1\farcs1) of the masers at the ``J2'' northeastern end, while their average speed
%(38~\kms) is about the double. Then, 
%Estimating the jet momentum rate from maser positions and velocities, 
%the values determined using separately the ``J2'' southwestern and northeastern maser clusters are in good agreement, which
%supports our thesis that the {\em same} outflow
%is driving the motion of both ``J2'' maser lobes. 
The more scattered distribution of proper motions of the water masers at the ``J2'' northeastern end could reflect turbulence induced by the
heating and photoevaporation of the {\em northeastern} side of the HMC, which is facing the O-type star responsible
for the \ion{H}{II} region observed 5\arcsec \ to NE of the HMC.

The orientation (PA$\approx$70\degr) of the ``J2'' jet agrees well with that (PA = 68\degr) of the
\Vlsr\ gradient across the HMC revealed in several transitions of CH$_3$CN and also
in the $^{12}$CO(2-1) line by \citet{Ces11}. Besides, the ``J2'' jet is powerful enough to explain
the large momentum rate, 0.3~$M_{\odot}$~yr$^{-1}$~km~s$^{-1}$, estimated from 
CH$_{3}^{13}$CN by \citet{Ces11} under the hypothesis that
the \Vlsr\ gradient represents a collimated outflow. These two facts
lead us to favour the interpretation of the  CH$_3$CN \Vlsr\ gradient in terms
of a compact and collimated outflow. 
The proper motions of the CH$_3$OH 6.7~GHz masers, mostly diverging from the HMC center,
%and the \Vlsr\ distribution of the OH 1.6~GHz masers, elongated at close angle with the
%CH$_3$CN \Vlsr\ gradient and showing a similar change in \Vlsr\ with position, 
also witness the expansion of the HMC gas driven by the ``J1'' and ``J2'' jets.
%  is expanding, which is probably driven by the sistem of double 
%(almost perpendicular) jets traced by the water masers inside the HMC.

As already indicated by previous results for other MYSOs, 
this work on the \Gd\ HMC confirms that VLBI maser, 3-D kinematics can be relevant for the study
of massive star-formation. This technique is presently the only one which
enables to resolve the typical sizes (several 100~AU) of massive stellar clusters, 
and to study in 3-D the process of mass accretion and ejection around a
single MYSO.

\begin{acknowledgements}
    ``J.J. Li were supported by
the Chinese National Science Foundation, through grants NSF
11203082, NSF 11133008, NSF 11073054, BK2012494, and the Key
Laboratory for Radio Astronomy, CAS. 
A.Sanna acknowledges the financial support by the European Research Council for the ERC Advanced Grant GLOSTAR under contract no. 247078.''
\end{acknowledgements}

% for the bibliography, at the end
\bibliographystyle{aa} % style aa.bst
\bibliography{biblio} % your references Yourfile.bib

\begin{thebibliography}{52}
\expandafter\ifx\csname natexlab\endcsname\relax\def\natexlab#1{#1}\fi

\bibitem[{{Araya} {et~al.}(2008){Araya}, {Hofner}, {Kurtz}, {Olmi}, \&
  {Linz}}]{Ara08}
{Araya}, E., {Hofner}, P., {Kurtz}, S., {Olmi}, L., \& {Linz}, H. 2008, \apj,
  675, 420

\bibitem[{{Bate} {et~al.}(2000){Bate}, {Bonnell}, {Clarke}, {Lubow}, {Ogilvie},
  {Pringle}, \& {Tout}}]{Bat00}
{Bate}, M.~R., {Bonnell}, I.~A., {Clarke}, C.~J., {et~al.} 2000, \mnras, 317,
  773

\bibitem[{{Beltr{\' a}n} {et~al.}(2004){Beltr{\' a}n}, {Cesaroni}, {Neri},
  {Codella}, {Furuya}, {Testi}, \& {Olmi}}]{Bel04}
{Beltr{\' a}n}, M.~T., {Cesaroni}, R., {Neri}, R., {et~al.} 2004, \apjl, 601,
  L187

\bibitem[{{Beltr{\'a}n} {et~al.}(2011){Beltr{\'a}n}, {Cesaroni}, {Neri}, \&
  {Codella}}]{Bel11}
{Beltr{\'a}n}, M.~T., {Cesaroni}, R., {Neri}, R., \& {Codella}, C. 2011, \aap,
  525, A151+

\bibitem[{{Beltr{\'a}n} {et~al.}(2005){Beltr{\'a}n}, {Cesaroni}, {Neri},
  {Codella}, {Furuya}, {Testi}, \& {Olmi}}]{Bel05}
{Beltr{\'a}n}, M.~T., {Cesaroni}, R., {Neri}, R., {et~al.} 2005, \aap, 435, 901

\bibitem[{{Beuther} {et~al.}(2002){Beuther}, {Schilke}, {Sridharan}, {Menten},
  {Walmsley}, \& {Wyrowski}}]{Beu02b}
{Beuther}, H., {Schilke}, P., {Sridharan}, T.~K., {et~al.} 2002, \aap, 383, 892

\bibitem[{{Bonnell} \& {Bate}(2006)}]{Bon06}
{Bonnell}, I.~A. \& {Bate}, M.~R. 2006, \mnras, 370, 488

\bibitem[{{Bontemps} {et~al.}(1996){Bontemps}, {Andre}, {Terebey}, \&
  {Cabrit}}]{Bon96}
{Bontemps}, S., {Andre}, P., {Terebey}, S., \& {Cabrit}, S. 1996, \aap, 311,
  858

\bibitem[{{Brunthaler} {et~al.}(2011){Brunthaler}, {Reid}, {Menten}, {Zheng},
  {Bartkiewicz}, {Choi}, {Dame}, {Hachisuka}, {Immer}, {Moellenbrock},
  {Moscadelli}, {Rygl}, {Sanna}, {Sato}, {Wu}, {Xu}, \& {Zhang}}]{Bru11}
{Brunthaler}, A., {Reid}, M.~J., {Menten}, K.~M., {et~al.} 2011, Astronomische
  Nachrichten, 332, 461

\bibitem[{{Cesaroni} {et~al.}(2011){Cesaroni}, {Beltr{\'a}n}, {Zhang},
  {Beuther}, \& {Fallscheer}}]{Ces11}
{Cesaroni}, R., {Beltr{\'a}n}, M.~T., {Zhang}, Q., {Beuther}, H., \&
  {Fallscheer}, C. 2011, \aap, 533, A73

\bibitem[{{Cesaroni} {et~al.}(1994{\natexlab{a}}){Cesaroni}, {Churchwell},
  {Hofner}, {Walmsley}, \& {Kurtz}}]{Ces94b}
{Cesaroni}, R., {Churchwell}, E., {Hofner}, P., {Walmsley}, C.~M., \& {Kurtz},
  S. 1994{\natexlab{a}}, \aap, 288, 903

\bibitem[{{Cesaroni} {et~al.}(2007){Cesaroni}, {Galli}, {Lodato}, {Walmsley},
  \& {Zhang}}]{Ces07}
{Cesaroni}, R., {Galli}, D., {Lodato}, G., {Walmsley}, C.~M., \& {Zhang}, Q.
  2007, Protostars and Planets V, 197

\bibitem[{{Cesaroni} {et~al.}(2010){Cesaroni}, {Hofner}, {Araya}, \&
  {Kurtz}}]{Ces10}
{Cesaroni}, R., {Hofner}, P., {Araya}, E., \& {Kurtz}, S. 2010, \aap, 509, A50

\bibitem[{{Cesaroni} {et~al.}(1998){Cesaroni}, {Hofner}, {Walmsley}, \&
  {Churchwell}}]{Ces98}
{Cesaroni}, R., {Hofner}, P., {Walmsley}, C.~M., \& {Churchwell}, E. 1998,
  \aap, 331, 709

\bibitem[{{Cesaroni} {et~al.}(1994{\natexlab{b}}){Cesaroni}, {Olmi},
  {Walmsley}, {Churchwell}, \& {Hofner}}]{Ces94}
{Cesaroni}, R., {Olmi}, L., {Walmsley}, C.~M., {Churchwell}, E., \& {Hofner},
  P. 1994{\natexlab{b}}, \apjl, 435, L137

\bibitem[{{Charnley} {et~al.}(1992){Charnley}, {Tielens}, \& {Millar}}]{Cha92}
{Charnley}, S.~B., {Tielens}, A.~G.~G.~M., \& {Millar}, T.~J. 1992, \apjl, 399,
  L71

\bibitem[{{Cragg} {et~al.}(2002){Cragg}, {Sobolev}, \& {Godfrey}}]{Cra02}
{Cragg}, D.~M., {Sobolev}, A.~M., \& {Godfrey}, P.~D. 2002, \mnras, 331, 521

\bibitem[{{Cunningham} {et~al.}(2011){Cunningham}, {Klein}, {Krumholz}, \&
  {McKee}}]{Cun11}
{Cunningham}, A.~J., {Klein}, R.~I., {Krumholz}, M.~R., \& {McKee}, C.~F. 2011,
  \apj, 740, 107

\bibitem[{{De Buizer}(2003)}]{DeB03}
{De Buizer}, J.~M. 2003, \mnras, 341, 277

\bibitem[{{De Buizer} {et~al.}(2009){De Buizer}, {Redman}, {Longmore},
  {Caswell}, \& {Feldman}}]{DeB09}
{De Buizer}, J.~M., {Redman}, R.~O., {Longmore}, S.~N., {Caswell}, J., \&
  {Feldman}, P.~A. 2009, \aap, 493, 127

\bibitem[{{Elitzur} {et~al.}(1989){Elitzur}, {Hollenbach}, \& {McKee}}]{eli89}
{Elitzur}, M., {Hollenbach}, D.~J., \& {McKee}, C.~F. 1989, \apj, 346, 983

\bibitem[{{Garrod}(2008)}]{Gar08}
{Garrod}, R.~T. 2008, \aap, 491, 239

\bibitem[{{Garrod} \& {Herbst}(2006)}]{Gar06}
{Garrod}, R.~T. \& {Herbst}, E. 2006, \aap, 457, 927

\bibitem[{{Girart} {et~al.}(2009){Girart}, {Beltr{\'a}n}, {Zhang}, {Rao}, \&
  {Estalella}}]{Gir09}
{Girart}, J.~M., {Beltr{\'a}n}, M.~T., {Zhang}, Q., {Rao}, R., \& {Estalella},
  R. 2009, Science, 324, 1408

\bibitem[{{Goddi} {et~al.}(2011){Goddi}, {Moscadelli}, \& {Sanna}}]{God11}
{Goddi}, C., {Moscadelli}, L., \& {Sanna}, A. 2011, \aap, 535, L8

\bibitem[{{Hartigan} {et~al.}(2011){Hartigan}, {Frank}, {Foster}, {Wilde},
  {Douglas}, {Rosen}, {Coker}, {Blue}, \& {Hansen}}]{Har11}
{Hartigan}, P., {Frank}, A., {Foster}, J.~M., {et~al.} 2011, \apj, 736, 29

\bibitem[{{Henney} {et~al.}(2009){Henney}, {Arthur}, {de Colle}, \&
  {Mellema}}]{Hen09}
{Henney}, W.~J., {Arthur}, S.~J., {de Colle}, F., \& {Mellema}, G. 2009,
  \mnras, 398, 157

\bibitem[{{Hofner} {et~al.}(2007){Hofner}, {Cesaroni}, {Olmi},
  {Rodr{\'{\i}}guez}, {Mart{\'{\i}}}, \& {Araya}}]{Hof07}
{Hofner}, P., {Cesaroni}, R., {Olmi}, L., {et~al.} 2007, \aap, 465, 197

\bibitem[{{K{\" o}nigl}(1999)}]{Kon99}
{K{\" o}nigl}, A. 1999, New Astronomy Review, 43, 67

\bibitem[{{Krumholz} {et~al.}(2009){Krumholz}, {Klein}, {McKee}, {Offner}, \&
  {Cunningham}}]{Kru09}
{Krumholz}, M.~R., {Klein}, R.~I., {McKee}, C.~F., {Offner}, S.~S.~R., \&
  {Cunningham}, A.~J. 2009, Science, 323, 754

\bibitem[{{Kuiper} {et~al.}(2010){Kuiper}, {Klahr}, {Beuther}, \&
  {Henning}}]{Kui10}
{Kuiper}, R., {Klahr}, H., {Beuther}, H., \& {Henning}, T. 2010, \apj, 722,
  1556

\bibitem[{{Kuiper} {et~al.}(2011){Kuiper}, {Klahr}, {Beuther}, \&
  {Henning}}]{Kui11}
{Kuiper}, R., {Klahr}, H., {Beuther}, H., \& {Henning}, T. 2011, \apj, 732, 20

\bibitem[{{Li} {et~al.}(2012){Li}, {Moscadelli}, {Cesaroni}, {Furuya}, {Xu},
  {Usuda}, {Menten}, {Pestalozzi}, {Elia}, \& {Schisano}}]{Li12}
{Li}, J.~J., {Moscadelli}, L., {Cesaroni}, R., {et~al.} 2012, \apj, 749, 47

\bibitem[{{L{\'o}pez-Sepulcre} {et~al.}(2009){L{\'o}pez-Sepulcre}, {Codella},
  {Cesaroni}, {Marcelino}, \& {Walmsley}}]{Lop09}
{L{\'o}pez-Sepulcre}, A., {Codella}, C., {Cesaroni}, R., {Marcelino}, N., \&
  {Walmsley}, C.~M. 2009, \aap, 499, 811

\bibitem[{{Marti} {et~al.}(1998){Marti}, {Rodriguez}, \& {Reipurth}}]{Mar98}
{Marti}, J., {Rodriguez}, L.~F., \& {Reipurth}, B. 1998, \apj, 502, 337

\bibitem[{{Moscadelli} {et~al.}(2011){Moscadelli}, {Cesaroni}, {Rioja},
  {Dodson}, \& {Reid}}]{Mos11}
{Moscadelli}, L., {Cesaroni}, R., {Rioja}, M.~J., {Dodson}, R., \& {Reid},
  M.~J. 2011, \aap, 526, A66+

\bibitem[{{Moscadelli} {et~al.}(2007){Moscadelli}, {Goddi}, {Cesaroni},
  {Beltr{\'a}n}, \& {Furuya}}]{Mos07}
{Moscadelli}, L., {Goddi}, C., {Cesaroni}, R., {Beltr{\'a}n}, M.~T., \&
  {Furuya}, R.~S. 2007, \aap, 472, 867

\bibitem[{{Moscadelli} {et~al.}(2010){Moscadelli}, {Xu}, \& {Chen}}]{Mos10}
{Moscadelli}, L., {Xu}, Y., \& {Chen}, X. 2010, \apj, 716, 1356

\bibitem[{{Osorio} {et~al.}(2009){Osorio}, {Anglada}, {Lizano}, \&
  {D'Alessio}}]{Oso09}
{Osorio}, M., {Anglada}, G., {Lizano}, S., \& {D'Alessio}, P. 2009, \apj, 694,
  29

\bibitem[{{Reipurth} {et~al.}(1999){Reipurth}, {Rodr{\'{\i}}guez}, \&
  {Chini}}]{Rei99}
{Reipurth}, B., {Rodr{\'{\i}}guez}, L.~F., \& {Chini}, R. 1999, \aj, 118, 983

\bibitem[{{Rodr{\'{\i}}guez} {et~al.}(2008){Rodr{\'{\i}}guez}, {Moran},
  {Franco-Hern{\'a}ndez}, {Garay}, {Brooks}, \& {Mardones}}]{Rod08}
{Rodr{\'{\i}}guez}, L.~F., {Moran}, J.~M., {Franco-Hern{\'a}ndez}, R., {et~al.}
  2008, \aj, 135, 2370

\bibitem[{{Sanna} {et~al.}(2010{\natexlab{a}}){Sanna}, {Moscadelli},
  {Cesaroni}, {Tarchi}, {Furuya}, \& {Goddi}}]{San10a}
{Sanna}, A., {Moscadelli}, L., {Cesaroni}, R., {et~al.} 2010{\natexlab{a}},
  \aap, 517, A71+

\bibitem[{{Sanna} {et~al.}(2010{\natexlab{b}}){Sanna}, {Moscadelli},
  {Cesaroni}, {Tarchi}, {Furuya}, \& {Goddi}}]{San10b}
{Sanna}, A., {Moscadelli}, L., {Cesaroni}, R., {et~al.} 2010{\natexlab{b}},
  \aap, 517, A78+

\bibitem[{{Sch{\"o}nrich} {et~al.}(2010){Sch{\"o}nrich}, {Binney}, \&
  {Dehnen}}]{Sch10}
{Sch{\"o}nrich}, R., {Binney}, J., \& {Dehnen}, W. 2010, \mnras, 403, 1829

\bibitem[{{Seifried} {et~al.}(2011){Seifried}, {Banerjee}, {Klessen}, {Duffin},
  \& {Pudritz}}]{Sei11}
{Seifried}, D., {Banerjee}, R., {Klessen}, R.~S., {Duffin}, D., \& {Pudritz},
  R.~E. 2011, \mnras, 417, 1054

\bibitem[{{Shepherd} {et~al.}(2000){Shepherd}, {Yu}, {Bally}, \&
  {Testi}}]{She00}
{Shepherd}, D.~S., {Yu}, K.~C., {Bally}, J., \& {Testi}, L. 2000, \apj, 535,
  833

\bibitem[{{Terquem} {et~al.}(1999){Terquem}, {Eisl{\"o}ffel}, {Papaloizou}, \&
  {Nelson}}]{Ter99}
{Terquem}, C., {Eisl{\"o}ffel}, J., {Papaloizou}, J.~C.~B., \& {Nelson}, R.~P.
  1999, \apjl, 512, L131

\bibitem[{{Vaidya} {et~al.}(2011){Vaidya}, {Fendt}, {Beuther}, \&
  {Porth}}]{Vai11}
{Vaidya}, B., {Fendt}, C., {Beuther}, H., \& {Porth}, O. 2011, \apj, 742, 56

\bibitem[{{Wu} {et~al.}(2005){Wu}, {Zhang}, {Chen}, {Yang}, {Wei}, \&
  {Ho}}]{Wu05}
{Wu}, Y., {Zhang}, Q., {Chen}, H., {et~al.} 2005, \aj, 129, 330

\bibitem[{{Yorke}(2002)}]{Yor02b}
{Yorke}, H.~W. 2002, in Astronomical Society of the Pacific Conference Series,
  Vol. 267, Hot Star Workshop III: The Earliest Phases of Massive Star Birth,
  ed. P.~{Crowther}, 165

\bibitem[{{Yorke} \& {Sonnhalter}(2002)}]{Yor02a}
{Yorke}, H.~W. \& {Sonnhalter}, C. 2002, \apj, 569, 846

\bibitem[{{Zhang} {et~al.}(2005){Zhang}, {Hunter}, {Brand}, {Sridharan},
  {Cesaroni}, {Molinari}, {Wang}, \& {Kramer}}]{Zha05}
{Zhang}, Q., {Hunter}, T.~R., {Brand}, J., {et~al.} 2005, \apj, 625, 864

\end{thebibliography}

\longtab{4}{
\begin{longtable}{rrrrrrrr}
\caption{\label{h2o_tab} 22~GHz H$_2$O Maser Parameters} \\
\hline\hline
%!comp  found_epoch   int_av(Jy|beam)  vel_av(km|s)   RA(mas) +/$-$ e_RA(mas)     DEC(mas) +/$-$ e_DEC(mas)      vx(km|s) +/$-$ e_vx(km|s)      vy(km|s) +/$-$ e_vy(km|s)  
\multicolumn{1}{c}{Feature} & \multicolumn{1}{c}{Epochs of} & \multicolumn{1}{c}{I$_{\rm peak}$} & \multicolumn{1}{c}{$V_{\rm LSR}$} & \multicolumn{1}{c}{$\Delta~x$} & \multicolumn{1}{c}{$\Delta~y$} & \multicolumn{1}{c}{$V_{x}$} & \multicolumn{1}{c}{$V_{y}$} \\
\multicolumn{1}{c}{Number}  & \multicolumn{1}{c}{Detection} & \multicolumn{1}{c}{(Jy beam$^{-1}$)} & \multicolumn{1}{c}{(km s$^{-1}$)} & \multicolumn{1}{c}{(mas)} & \multicolumn{1}{c}{(mas)} & \multicolumn{1}{c}{(km s$^{-1}$)} & \multicolumn{1}{c}{(km s$^{-1}$)} \\
\hline
\endfirsthead
\caption{continued.}\\
\hline\hline
\multicolumn{1}{c}{Feature} & \multicolumn{1}{c}{Epochs of} & \multicolumn{1}{c}{I$_{\rm peak}$} & \multicolumn{1}{c}{$V_{\rm LSR}$} & \multicolumn{1}{c}{$\Delta~x$} & \multicolumn{1}{c}{$\Delta~y$} & \multicolumn{1}{c}{$V_{x}$} & \multicolumn{1}{c}{$V_{y}$} \\
\multicolumn{1}{c}{Number}  & \multicolumn{1}{c}{Detection} & \multicolumn{1}{c}{(Jy beam$^{-1}$)} & \multicolumn{1}{c}{(km s$^{-1}$)} & \multicolumn{1}{c}{(mas)} & \multicolumn{1}{c}{(mas)} & \multicolumn{1}{c}{(km s$^{-1}$)} & \multicolumn{1}{c}{(km s$^{-1}$)} \\
\hline
\endhead
\hline
\endfoot
%!
%! feature 103 e 112 have not reliable (and not plotted) proper motions, even if they are observed at three epochs
%!
    0 &        1,2,3,4 &       ... &    98.4 &   173.65$\pm$0.06 &   767.21$\pm$0.06 & 0.0$\pm$0.0  & 0.0$\pm$0.0  \\
    1 &        1,2,3,4 &    108.04 &    98.0 &     0.00$\pm$0.00  &     0.00$\pm$0.00  &    0.6$\pm$5.7 &  $$-$$30.7$\pm$5.9  \\
    2 &        1,2,3,4 &     30.13 &    97.5 &   760.41$\pm$0.08 &   211.81$\pm$0.08 &  $-$13.3$\pm$5.7 &    1.3$\pm$5.8  \\
    3 &        1,2,3,4 &     21.02 &    97.2 &  $-$625.59$\pm$0.08 &   390.26$\pm$0.08 &  $-$12.4$\pm$5.4 &   $-$6.4$\pm$5.5  \\
    4 &        1,2,3,4 &     20.30 &   100.9 &   $-$16.62$\pm$0.07 &    $-$5.71$\pm$0.08 &    6.4$\pm$5.5 &  $-$36.0$\pm$5.7  \\
    5 &        1,2,3,4 &     12.75 &    95.2 &  $-$514.10$\pm$0.08 &   280.57$\pm$0.08 &  $-$24.9$\pm$5.6 &  $-$29.1$\pm$5.7  \\
    6 &              1 &     11.49 &    97.8 &   759.52$\pm$0.07 &   211.05$\pm$0.08 &    ... &    ...  \\
    7 &          1,2,3 &      8.99 &    94.8 &   123.56$\pm$0.08 &   476.60$\pm$0.08 &   36.5$\pm$9.1 &    1.9$\pm$9.5  \\
    8 &        1,2,3,4 &      6.71 &   100.1 &   827.44$\pm$0.07 &  1418.08$\pm$0.08 &   16.0$\pm$5.5 &    0.1$\pm$5.6  \\
    9 &        1,2,3,4 &      6.45 &    93.6 &  $-$514.71$\pm$0.08 &   247.81$\pm$0.08 &  $-$21.8$\pm$5.6 &   $-$7.2$\pm$5.7  \\
   10 &        1,2,3,4 &      6.41 &    97.0 &   772.36$\pm$0.08 &   245.23$\pm$0.08 &   $-$2.1$\pm$5.5 &   26.3$\pm$5.7  \\
   11 &        1,2,3,4 &      6.37 &    99.5 &    15.45$\pm$0.08 &    14.27$\pm$0.08 &   $-$3.1$\pm$5.5 &  $-$18.6$\pm$5.6  \\
   12 &        1,2,3,4 &      5.83 &   102.1 &   860.39$\pm$0.08 &  1371.82$\pm$0.08 &    3.7$\pm$5.5 &   19.4$\pm$5.6  \\
   13 &        1,2,3,4 &      5.51 &    98.3 &    $-$2.99$\pm$0.07 &     0.20$\pm$0.08 &   65.6$\pm$5.5 &  $-$21.4$\pm$5.8  \\
   14 &        1,2,3,4 &      4.95 &    96.0 &  $-$513.87$\pm$0.07 &   282.77$\pm$0.08 &  $-$20.5$\pm$5.4 &  $-$12.4$\pm$5.5  \\
   15 &              3 &      4.71 &    96.0 &   $-$12.21$\pm$0.07 &    $-$3.91$\pm$0.08 &    ... &    ...  \\
   16 &            1,3 &      4.67 &    97.5 &   761.64$\pm$0.07 &   213.02$\pm$0.08 &    ... &    ...  \\
   17 &        1,2,3,4 &      4.66 &   100.6 &   596.73$\pm$0.08 &  1444.51$\pm$0.08 &   $-$4.1$\pm$5.7 &   15.0$\pm$5.9  \\
   18 &        1,2,3,4 &      4.28 &    99.8 &   823.47$\pm$0.07 &  1424.87$\pm$0.08 &    5.6$\pm$5.5 &   $-$0.7$\pm$5.6  \\
   19 &              3 &      4.27 &    87.4 &   106.77$\pm$0.07 &   458.18$\pm$0.08 &    ... &    ...  \\
   20 &        1,2,3,4 &      4.20 &   113.2 &  $-$106.13$\pm$0.08 &   432.19$\pm$0.08 &   28.9$\pm$5.8 &  $-$60.0$\pm$6.0  \\
   21 &        1,2,3,4 &      4.19 &    98.5 &    10.63$\pm$0.08 &     5.19$\pm$0.08 &   40.5$\pm$5.5 &  $-$28.3$\pm$5.7  \\
   22 &        1,2,3,4 &      3.89 &    97.2 &   771.83$\pm$0.07 &   239.24$\pm$0.08 &    1.3$\pm$5.4 &   26.6$\pm$5.7  \\
   23 &        1,2,3,4 &      3.77 &    99.3 &   884.14$\pm$0.07 &  1345.64$\pm$0.08 &   14.7$\pm$5.5 &   18.7$\pm$5.6  \\
   24 &        1,2,3,4 &      3.72 &    97.6 &   771.01$\pm$0.08 &   234.36$\pm$0.08 &   $-$2.5$\pm$5.5 &   23.6$\pm$5.7  \\
   25 &          1,2,3 &      3.13 &   118.4 &    41.31$\pm$0.08 &  1187.73$\pm$0.08 &   $-$9.7$\pm$9.0 &   82.7$\pm$9.2  \\
   26 &        1,2,3,4 &      3.12 &    99.3 &   881.46$\pm$0.08 &  1326.57$\pm$0.08 &   10.0$\pm$5.5 &   11.5$\pm$5.8  \\
   27 &              3 &      2.43 &    93.1 &   $-$12.38$\pm$0.07 &    $-$3.67$\pm$0.08 &    ... &    ...  \\
   28 &        1,2,3,4 &      2.30 &   100.2 &   207.47$\pm$0.08 &  1228.88$\pm$0.08 &   10.3$\pm$5.5 &   $-$4.4$\pm$5.8  \\
   29 &          2,3,4 &      2.28 &    97.2 &   770.52$\pm$0.08 &   232.86$\pm$0.08 &   15.8$\pm$8.4 &   66.7$\pm$9.2  \\
   30 &        1,2,3,4 &      2.28 &    89.5 &   111.09$\pm$0.08 &   461.95$\pm$0.08 &   32.9$\pm$5.6 &  $-$50.2$\pm$5.8  \\
   31 &              2 &      1.89 &    99.5 &   $-$11.96$\pm$0.08 &    $-$3.17$\pm$0.08 &    ... &    ...  \\
   32 &        1,2,3,4 &      1.81 &   107.8 &  $-$401.49$\pm$0.08 &    87.41$\pm$0.08 &    3.8$\pm$6.5 &  $-$79.9$\pm$7.1  \\
   33 &        1,2,3,4 &      1.74 &   100.7 &   207.63$\pm$0.08 &  1235.46$\pm$0.08 &   20.8$\pm$5.6 &   10.6$\pm$5.9  \\
   34 &        1,2,3,4 &      1.67 &   101.2 &   $-$54.19$\pm$0.08 &  1052.12$\pm$0.08 &  $-$17.0$\pm$5.4 &   26.9$\pm$5.7  \\
   35 &          2,3,4 &      1.62 &   114.4 &    35.79$\pm$0.09 &  1184.51$\pm$0.11 &  $-$46.6$\pm$10.2 &   80.2$\pm$11.3  \\
   36 &              4 &      1.55 &    87.3 &   107.96$\pm$0.07 &   458.62$\pm$0.07 &    ... &    ...  \\
   37 &        1,2,3,4 &      1.53 &    95.8 &  $-$128.10$\pm$0.08 &   203.66$\pm$0.08 &  $-$24.2$\pm$5.6 &   $-$4.2$\pm$5.8  \\
   38 &              1 &      1.46 &    95.9 &   $-$12.34$\pm$0.08 &    $-$3.66$\pm$0.08 &    ... &    ...  \\
   39 &              1 &      1.46 &    98.4 &     0.74$\pm$0.07 &    $-$0.23$\pm$0.08 &    ... &    ...  \\
   40 &              1 &      1.43 &    98.7 &   $-$11.99$\pm$0.08 &    $-$3.22$\pm$0.08 &    ... &    ...  \\
   41 &          2,3,4 &      1.42 &   118.5 &    24.08$\pm$0.08 &  1048.78$\pm$0.08 &  $-$12.1$\pm$8.7 &   45.0$\pm$8.9  \\
   42 &              2 &      1.40 &    96.7 &   $-$12.26$\pm$0.08 &    $-$3.80$\pm$0.08 &    ... &    ...  \\
   43 &          1,2,3 &      1.36 &    96.9 &   148.53$\pm$0.07 &  1412.51$\pm$0.08 &    8.8$\pm$10.4 &    1.1$\pm$11.7  \\
   44 &        1,2,3,4 &      1.31 &    99.6 &   $-$53.87$\pm$0.08 &  1057.11$\pm$0.08 &  $-$16.3$\pm$5.7 &   16.6$\pm$6.1  \\
   45 &              4 &      1.30 &    96.3 &   $-$12.19$\pm$0.08 &    $-$4.02$\pm$0.08 &    ... &    ...  \\
   46 &        1,2,3,4 &      1.29 &    94.1 &    $-$8.91$\pm$0.08 &  1202.27$\pm$0.08 &   $-$7.6$\pm$5.6 &   11.5$\pm$5.8  \\
   47 &        1,2,3,4 &      1.21 &    99.3 &   970.34$\pm$0.07 &  1068.50$\pm$0.08 &    8.4$\pm$6.1 &   29.7$\pm$6.7  \\
   48 &        1,2,3,4 &      1.19 &    95.1 &  $-$514.60$\pm$0.08 &   275.71$\pm$0.08 &  $-$33.8$\pm$5.5 &  $-$56.3$\pm$5.8  \\
   49 &              1 &      0.99 &    96.6 &    15.87$\pm$0.08 &     8.46$\pm$0.08 &    ... &    ...  \\
   50 &        1,2,3,4 &      0.92 &    95.2 &   135.06$\pm$0.08 &   516.27$\pm$0.08 &   36.6$\pm$5.7 &    4.6$\pm$6.2  \\
   51 &              4 &      0.77 &    94.9 &   970.81$\pm$0.07 &  1069.10$\pm$0.08 &    ... &    ...  \\
   52 &              3 &      0.75 &    94.9 &  $-$514.38$\pm$0.08 &   277.10$\pm$0.08 &    ... &    ...  \\
   53 &        1,2,3,4 &      0.70 &   101.8 &    10.05$\pm$0.08 &  1206.54$\pm$0.08 &  $-$30.2$\pm$5.9 &   20.6$\pm$6.4  \\
   54 &        1,2,3,4 &      0.70 &    96.7 &   103.78$\pm$0.08 &   472.25$\pm$0.08 &   44.6$\pm$5.9 &   14.6$\pm$6.1  \\
   55 &        1,2,3,4 &      0.70 &   101.4 &   615.77$\pm$0.08 &  1450.81$\pm$0.08 &   $-$8.6$\pm$5.7 &   17.2$\pm$6.0  \\
   56 &        1,2,3,4 &      0.69 &    87.3 &   110.86$\pm$0.08 &   461.48$\pm$0.09 &   43.9$\pm$5.8 &  $-$24.6$\pm$6.5  \\
   57 &              4 &      0.65 &    99.7 &   $-$17.09$\pm$0.08 &    $-$7.96$\pm$0.08 &    ... &    ...  \\
   58 &        1,2,3,4 &      0.63 &    96.9 &   159.05$\pm$0.08 &  1395.47$\pm$0.09 &    2.5$\pm$6.1 &   62.3$\pm$6.7  \\
   59 &              4 &      0.61 &    92.5 &   $-$12.36$\pm$0.08 &    $-$3.79$\pm$0.08 &    ... &    ...  \\
   60 &        1,2,3,4 &      0.59 &    96.6 &    82.75$\pm$0.08 &   463.46$\pm$0.09 &   $-$3.7$\pm$6.1 &   $-$9.3$\pm$6.7  \\
   61 &              1 &      0.56 &   101.1 &  $-$405.27$\pm$0.08 &    89.91$\pm$0.08 &    ... &    ...  \\
   62 &          2,3,4 &      0.55 &   103.5 &  $-$406.59$\pm$0.08 &    90.46$\pm$0.09 &  $-$90.0$\pm$9.1 &  $-$75.7$\pm$9.7  \\
   63 &        1,2,3,4 &      0.54 &    99.6 &   878.77$\pm$0.08 &  1316.82$\pm$0.08 &    2.9$\pm$6.2 &    5.7$\pm$7.4  \\
   64 &              4 &      0.54 &    98.8 &   200.87$\pm$0.08 &  1309.50$\pm$0.08 &    ... &    ...  \\
   65 &        1,2,3,4 &      0.52 &   103.6 &   $-$41.44$\pm$0.08 &  1125.30$\pm$0.08 &  $-$21.4$\pm$5.8 &   15.6$\pm$6.2  \\
   66 &              3 &      0.51 &    98.8 &   200.78$\pm$0.08 &  1309.34$\pm$0.08 &    ... &    ...  \\
   67 &              2 &      0.50 &    99.9 &  $-$404.50$\pm$0.09 &    89.78$\pm$0.09 &    ... &    ...  \\
   68 &              4 &      0.46 &    99.3 &    15.55$\pm$0.08 &     8.47$\pm$0.08 &    ... &    ...  \\
   69 &              4 &      0.43 &    98.5 &   202.48$\pm$0.08 &  1300.13$\pm$0.09 &    ... &    ...  \\
   70 &        1,2,3,4 &      0.42 &    94.7 &   103.76$\pm$0.08 &  1349.41$\pm$0.09 &  $-$19.2$\pm$5.9 &   19.6$\pm$6.6  \\
   71 &            1,2 &      0.42 &   100.5 &   590.55$\pm$0.08 &  1446.26$\pm$0.09 &    ... &    ...  \\
   72 &        1,2,3,4 &      0.41 &   101.4 &   591.27$\pm$0.08 &  1438.58$\pm$0.08 &  $-$19.0$\pm$5.8 &   20.6$\pm$6.2  \\
   73 &              4 &      0.41 &    99.3 &   883.47$\pm$0.08 &  1332.41$\pm$0.09 &    ... &    ...  \\
   74 &              4 &      0.41 &    95.0 &   124.38$\pm$0.08 &   477.45$\pm$0.08 &    ... &    ...  \\
   75 &              1 &      0.41 &   127.7 &  $-$108.53$\pm$0.08 &   430.49$\pm$0.08 &    ... &    ...  \\
   76 &            1,2 &      0.41 &   100.3 &    12.37$\pm$0.08 &    14.79$\pm$0.09 &    ... &    ...  \\
   77 &              1 &      0.41 &    95.7 &    62.90$\pm$0.08 &  1271.66$\pm$0.08 &    ... &    ...  \\
   78 &        1,2,3,4 &      0.40 &   101.8 &   242.22$\pm$0.08 &  1240.44$\pm$0.10 &   12.8$\pm$6.2 &   18.9$\pm$7.3  \\
   79 &          1,2,3 &      0.39 &    99.6 &   881.68$\pm$0.08 &  1354.22$\pm$0.09 &    1.5$\pm$11.6 &   14.7$\pm$15.5  \\
   80 &        1,2,3,4 &      0.37 &   102.3 &   623.19$\pm$0.08 &  1447.14$\pm$0.08 &   $-$7.7$\pm$6.1 &   13.7$\pm$6.5  \\
   81 &              1 &      0.37 &    99.9 &   871.98$\pm$0.08 &  1307.43$\pm$0.10 &    ... &    ...  \\
   82 &              4 &      0.36 &    93.0 &   124.00$\pm$0.08 &   476.56$\pm$0.09 &    ... &    ...  \\
   83 &          1,2,4 &      0.36 &   100.7 &    83.42$\pm$0.08 &   274.56$\pm$0.10 &   15.4$\pm$6.8 &  $-$11.1$\pm$8.1  \\
   84 &            2,4 &      0.35 &   100.5 &   822.73$\pm$0.08 &  1426.15$\pm$0.10 &    ... &    ...  \\
   85 &        1,2,3,4 &      0.34 &    96.7 &  $-$623.32$\pm$0.08 &   376.77$\pm$0.09 &  $-$13.0$\pm$6.6 &    0.0$\pm$7.5  \\
   86 &              4 &      0.34 &    99.5 &   204.01$\pm$0.08 &  1268.80$\pm$0.09 &    ... &    ...  \\
   87 &              1 &      0.34 &   101.4 &   $-$12.45$\pm$0.08 &    $-$3.26$\pm$0.08 &    ... &    ...  \\
   88 &              4 &      0.33 &   100.0 &    29.61$\pm$0.09 &    12.33$\pm$0.10 &    ... &    ...  \\
   89 &              4 &      0.33 &    99.7 &   203.21$\pm$0.09 &  1285.90$\pm$0.11 &    ... &    ...  \\
   90 &              4 &      0.33 &    98.6 &     8.88$\pm$0.08 &     4.77$\pm$0.09 &    ... &    ...  \\
   91 &        1,2,3,4 &      0.32 &   104.9 &   193.07$\pm$0.08 &   782.37$\pm$0.09 &   11.7$\pm$6.0 &   $-$4.4$\pm$6.7  \\
   92 &          2,3,4 &      0.31 &    95.0 &  $-$516.60$\pm$0.09 &   256.92$\pm$0.09 &  $-$12.9$\pm$9.6 &  $-$12.7$\pm$10.4  \\
   93 &              3 &      0.31 &    94.6 &   970.64$\pm$0.08 &  1068.73$\pm$0.08 &    ... &    ...  \\
   94 &          1,2,3 &      0.31 &   102.3 &   859.58$\pm$0.08 &  1412.65$\pm$0.09 &  $-$28.7$\pm$10.4 &   51.8$\pm$12.4  \\
   95 &              4 &      0.31 &    99.1 &   $-$85.10$\pm$0.08 &    96.72$\pm$0.09 &    ... &    ...  \\
   96 &          1,2,3 &      0.30 &   123.6 &   $-$91.29$\pm$0.08 &   448.82$\pm$0.08 &   $-$2.5$\pm$10.2 &  $-$39.5$\pm$11.2  \\
   97 &        1,2,3,4 &      0.30 &    95.7 &  $-$516.55$\pm$0.08 &   265.90$\pm$0.08 &  $-$23.0$\pm$5.9 &    3.6$\pm$7.0  \\
   98 &              3 &      0.30 &    99.3 &    15.38$\pm$0.09 &     8.48$\pm$0.09 &    ... &    ...  \\
   99 &          2,3,4 &      0.29 &    95.1 &  $-$616.35$\pm$0.08 &   433.17$\pm$0.09 &  $-$12.4$\pm$9.3 &   $-$9.3$\pm$10.3  \\
  100 &          1,2,3 &      0.28 &    95.8 &    47.97$\pm$0.08 &   194.52$\pm$0.08 &   14.8$\pm$10.1 &  $-$21.2$\pm$11.2  \\
  101 &              3 &      0.28 &    96.5 &   970.57$\pm$0.08 &  1068.80$\pm$0.09 &    ... &    ...  \\
  102 &              4 &      0.27 &    90.6 &   124.52$\pm$0.08 &   477.19$\pm$0.09 &    ... &    ...  \\
  103\tablefootmark{a} &          1,2,3 &      0.27 &    86.4 &   106.46$\pm$0.08 &   458.25$\pm$0.09 &    ... &    ...  \\
  104 &              4 &      0.27 &    99.3 &   $-$90.01$\pm$0.08 &   142.78$\pm$0.10 &    ... &    ...  \\
  105 &              4 &      0.26 &    98.6 &   871.68$\pm$0.09 &  1282.10$\pm$0.10 &    ... &    ...  \\
  106 &          1,3,4 &      0.25 &   101.5 &    10.67$\pm$0.08 &  1208.55$\pm$0.09 &  $-$32.1$\pm$6.3 &    3.6$\pm$7.4  \\
  107 &              3 &      0.24 &    99.5 &   203.91$\pm$0.09 &  1268.48$\pm$0.10 &    ... &    ...  \\
  108 &              2 &      0.24 &   100.1 &   $-$84.72$\pm$0.10 &    97.24$\pm$0.12 &    ... &    ...  \\
  109 &            2,3 &      0.24 &    95.7 &    53.11$\pm$0.09 &  1291.76$\pm$0.10 &    ... &    ...  \\
  110 &        1,2,3,4 &      0.24 &    96.4 &    92.11$\pm$0.08 &   467.69$\pm$0.09 &    8.3$\pm$6.3 &  $-$10.3$\pm$7.2  \\
  111 &            1,2 &      0.23 &   102.4 &   $-$94.09$\pm$0.08 &   164.25$\pm$0.08 &    ... &    ...  \\
  112\tablefootmark{a} &          1,2,3 &      0.22 &    87.4 &    41.52$\pm$0.08 &   525.37$\pm$0.09 &    ... &    ...  \\
  113 &              2 &      0.22 &   100.5 &   596.04$\pm$0.10 &  1485.14$\pm$0.14 &    ... &    ...  \\
  114 &              4 &      0.21 &   115.7 &    47.16$\pm$0.08 &  1065.65$\pm$0.09 &    ... &    ...  \\
  115 &              2 &      0.21 &    99.3 &    11.83$\pm$0.11 &   106.83$\pm$0.10 &    ... &    ...  \\
  116 &              1 &      0.20 &   101.7 &    96.99$\pm$0.08 &  1384.97$\pm$0.09 &    ... &    ...  \\
  117 &            2,3 &      0.20 &    95.1 &   135.26$\pm$0.09 &   558.19$\pm$0.09 &    ... &    ...  \\
  118 &              4 &      0.20 &    95.9 &   123.38$\pm$0.08 &   475.38$\pm$0.09 &    ... &    ...  \\
  119 &            3,4 &      0.19 &    95.7 &   121.26$\pm$0.08 &   475.30$\pm$0.09 &    ... &    ...  \\
  120 &          2,3,4 &      0.19 &    95.7 &    51.23$\pm$0.08 &  1291.83$\pm$0.09 &  $-$23.9$\pm$10.5 &   52.2$\pm$12.1  \\
  121 &              4 &      0.18 &    91.7 &   110.77$\pm$0.08 &   460.47$\pm$0.09 &    ... &    ...  \\
  122 &        1,2,3,4 &      0.18 &   103.6 &   $-$54.77$\pm$0.08 &  1049.45$\pm$0.09 &  $-$16.3$\pm$6.5 &   14.6$\pm$8.1  \\
  123 &        1,2,3,4 &      0.17 &    94.3 &   105.02$\pm$0.08 &  1351.10$\pm$0.10 &  $-$18.2$\pm$6.5 &   15.0$\pm$7.7  \\
  124 &            1,2 &      0.17 &   116.3 &    18.99$\pm$0.08 &  1162.29$\pm$0.09 &    ... &    ...  \\
  125 &              2 &      0.16 &   101.2 &   $-$17.73$\pm$0.14 &   $-$31.34$\pm$0.15 &    ... &    ...  \\
  126 &              1 &      0.16 &    92.3 &   120.38$\pm$0.08 &   469.82$\pm$0.09 &    ... &    ...  \\
  127 &              1 &      0.16 &   101.6 &   $-$53.24$\pm$0.08 &  1046.79$\pm$0.09 &    ... &    ...  \\
  128 &            1,3 &      0.16 &   101.7 &   194.02$\pm$0.09 &  1216.44$\pm$0.10 &    ... &    ...  \\
  129 &              1 &      0.15 &    91.0 &  $-$227.30$\pm$0.08 &   646.92$\pm$0.10 &    ... &    ...  \\
  130 &            3,4 &      0.14 &    91.7 &  $-$167.73$\pm$0.09 &   380.68$\pm$0.10 &    ... &    ...  \\
  131 &              4 &      0.14 &    84.2 &  $-$507.22$\pm$0.08 &   520.81$\pm$0.10 &    ... &    ...  \\
  132 &              4 &      0.14 &   111.2 &    20.79$\pm$0.08 &  1155.17$\pm$0.09 &    ... &    ...  \\
  133 &              2 &      0.14 &   106.7 &  $-$105.08$\pm$0.09 &   456.25$\pm$0.10 &    ... &    ...  \\
  134 &          1,2,3 &      0.14 &   105.0 &  $-$407.55$\pm$0.08 &    89.72$\pm$0.10 &  $-$89.4$\pm$11.3 &  $-$60.6$\pm$13.6  \\
  135 &              1 &      0.14 &    88.7 &   123.51$\pm$0.08 &   487.91$\pm$0.10 &    ... &    ...  \\
  136 &              1 &      0.14 &    94.0 &    $-$7.07$\pm$0.08 &  1203.64$\pm$0.09 &    ... &    ...  \\
  137 &              2 &      0.14 &   101.9 &   $-$17.06$\pm$0.10 &   $-$30.70$\pm$0.10 &    ... &    ...  \\
  138 &              3 &      0.13 &   115.0 &    42.78$\pm$0.08 &  1076.47$\pm$0.09 &    ... &    ...  \\
  139 &            1,2 &      0.13 &   104.5 &   $-$52.94$\pm$0.08 &  1062.76$\pm$0.10 &    ... &    ...  \\
  140 &              3 &      0.13 &    87.3 &   108.27$\pm$0.08 &   506.54$\pm$0.09 &    ... &    ...  \\
  141 &              2 &      0.13 &    96.5 &   131.24$\pm$0.10 &   488.19$\pm$0.11 &    ... &    ...  \\
  142 &            1,2 &      0.13 &    91.7 &   $-$92.49$\pm$0.08 &   677.89$\pm$0.09 &    ... &    ...  \\
  143 &        1,2,3,4 &      0.13 &    87.0 &   $-$62.91$\pm$0.08 &   279.80$\pm$0.10 &   22.6$\pm$8.0 &  $-$62.4$\pm$9.7  \\
  144 &        1,2,3,4 &      0.12 &   104.7 &   $-$45.49$\pm$0.08 &  1082.40$\pm$0.09 &  $-$27.6$\pm$7.4 &   $-$2.5$\pm$9.6  \\
  145 &              2 &      0.12 &   118.5 &    35.40$\pm$0.09 &  1184.50$\pm$0.10 &    ... &    ...  \\
  146 &            2,3 &      0.11 &   124.0 &    17.42$\pm$0.09 &  1158.01$\pm$0.11 &    ... &    ...  \\
  147 &              4 &      0.11 &    90.1 &   103.48$\pm$0.09 &   458.33$\pm$0.11 &    ... &    ...  \\
  148 &            3,4 &      0.11 &   103.4 &    33.25$\pm$0.10 &  1245.23$\pm$0.11 &    ... &    ...  \\
  149 &              2 &      0.11 &    93.7 &   $-$12.45$\pm$0.09 &    $-$3.75$\pm$0.11 &    ... &    ...  \\
  150 &              2 &      0.11 &    88.6 &   106.95$\pm$0.09 &   458.31$\pm$0.10 &    ... &    ...  \\
  151 &              2 &      0.11 &   117.4 &    50.55$\pm$0.10 &  1190.01$\pm$0.10 &    ... &    ...  \\
  152 &              1 &      0.11 &    96.3 &    $-$0.54$\pm$0.09 &  1199.17$\pm$0.10 &    ... &    ...  \\
  153 &              1 &      0.10 &   106.4 &   $-$95.93$\pm$0.08 &   163.34$\pm$0.10 &    ... &    ...  \\
  154 &              2 &      0.10 &   102.0 &   $-$53.00$\pm$0.10 &  1046.06$\pm$0.13 &    ... &    ...  \\
  155 &        1,2,3,4 &      0.10 &   104.3 &    30.90$\pm$0.08 &  1237.90$\pm$0.09 &  $-$14.9$\pm$7.6 &   16.7$\pm$9.2  \\
  156 &          1,2,3 &      0.10 &    92.1 &   $-$93.50$\pm$0.09 &   174.90$\pm$0.12 &  $-$23.3$\pm$12.0 &   $-$7.4$\pm$15.1  \\
  157 &            1,2 &      0.09 &    93.7 &  $-$109.34$\pm$0.08 &   170.88$\pm$0.10 &    ... &    ...  \\
  158 &              2 &      0.09 &   102.6 &   119.90$\pm$0.10 &   475.82$\pm$0.11 &    ... &    ...  \\
  159 &              4 &      0.09 &   115.4 &    22.03$\pm$0.10 &  1155.46$\pm$0.12 &    ... &    ...  \\
  160 &              1 &      0.09 &   103.5 &   597.29$\pm$0.09 &   987.29$\pm$0.11 &    ... &    ...  \\
  161 &            1,2 &      0.09 &   104.8 &   $-$56.70$\pm$0.09 &  1041.70$\pm$0.12 &    ... &    ...  \\
  162 &            1,2 &      0.09 &    92.7 &   167.30$\pm$0.08 &  1841.14$\pm$0.10 &    ... &    ...  \\
  163 &            2,4 &      0.09 &    87.9 &   106.06$\pm$0.10 &   457.91$\pm$0.11 &    ... &    ...  \\
  164 &              2 &      0.08 &    92.3 &   104.04$\pm$0.11 &   458.34$\pm$0.12 &    ... &    ...  \\
  165 &              1 &      0.08 &   112.0 &  $-$106.06$\pm$0.08 &   409.95$\pm$0.11 &    ... &    ...  \\
  166 &            1,2 &      0.08 &    91.6 &   151.34$\pm$0.08 &  1831.39$\pm$0.11 &    ... &    ...  \\
  167 &              3 &      0.07 &   102.9 &    25.19$\pm$0.12 &  1241.43$\pm$0.12 &    ... &    ...  \\
  168 &              1 &      0.07 &    92.9 &   $-$90.08$\pm$0.09 &   678.15$\pm$0.11 &    ... &    ...  \\
  169 &              1 &      0.06 &   113.2 &    $-$0.62$\pm$0.09 &  1063.01$\pm$0.11 &    ... &    ...  \\
  170 &              1 &      0.06 &    93.4 &   139.41$\pm$0.09 &   518.74$\pm$0.11 &    ... &    ...  \\
  171 &              1 &      0.05 &   111.8 &  $-$106.28$\pm$0.11 &   390.79$\pm$0.13 &    ... &    ...  \\
  172 &              1 &      0.05 &    88.7 &   103.29$\pm$0.11 &   458.90$\pm$0.12 &    ... &    ...  \\
  173 &              2 &      0.05 &   113.8 &     3.78$\pm$0.12 &  1128.38$\pm$0.15 &    ... &    ...  \\
\end{longtable}
\tablefoot{Column~1 gives the feature label number; column~2
lists the observing epochs at which the feature was detected;
columns~3~and~4 provide the intensity of the strongest spot
and the intensity-weighted LSR velocity, respectively, averaged over the
observing epochs; columns~5~and~6 give the position offsets (with
the associated errors) along the R.A. and Dec. axes, relative to the
feature~\#1, measured at the first epoch of detection; columns~7~and~8 give the components of the relative
proper motion (with the associated errors) along the R.A. and Dec.
axes, measured with respect to the reference feature~\#0 (the
``center of motion'').\\
\tablefoottext{a}{Feature with not reliable proper motion, even if observed at three epochs.}
}
}

\longtab{5}{
\begin{longtable}{rrrrrrrr}
\caption{\label{ch3oh_tab} 6.7~GHz CH$_3$OH Maser Parameters} \\
\hline\hline
%!comp  found_epoch   int_av(Jy|beam)  vel_av(km|s)   RA(mas) +/$-$ e_RA(mas)     DEC(mas) +/$-$ e_DEC(mas)      vx(km|s) +/$-$ e_vx(km|s)      vy(km|s) +/$-$ e_vy(km|s)  
\multicolumn{1}{c}{Feature} & \multicolumn{1}{c}{Epochs of} & \multicolumn{1}{c}{I$_{\rm peak}$} & \multicolumn{1}{c}{$V_{\rm LSR}$} & \multicolumn{1}{c}{$\Delta~x$} & \multicolumn{1}{c}{$\Delta~y$} & \multicolumn{1}{c}{$V_{x}$} & \multicolumn{1}{c}{$V_{y}$} \\
\multicolumn{1}{c}{Number}  & \multicolumn{1}{c}{Detection} & \multicolumn{1}{c}{(Jy beam$^{-1}$)} & \multicolumn{1}{c}{(km s$^{-1}$)} & \multicolumn{1}{c}{(mas)} & \multicolumn{1}{c}{(mas)} & \multicolumn{1}{c}{(km s$^{-1}$)} & \multicolumn{1}{c}{(km s$^{-1}$)} \\
\hline
\endfirsthead
\caption{continued.}\\
\hline\hline
\multicolumn{1}{c}{Feature} & \multicolumn{1}{c}{Epochs of} & \multicolumn{1}{c}{I$_{\rm peak}$} & \multicolumn{1}{c}{$V_{\rm LSR}$} & \multicolumn{1}{c}{$\Delta~x$} & \multicolumn{1}{c}{$\Delta~y$} & \multicolumn{1}{c}{$V_{x}$} & \multicolumn{1}{c}{$V_{y}$} \\
\multicolumn{1}{c}{Number}  & \multicolumn{1}{c}{Detection} & \multicolumn{1}{c}{(Jy beam$^{-1}$)} & \multicolumn{1}{c}{(km s$^{-1}$)} & \multicolumn{1}{c}{(mas)} & \multicolumn{1}{c}{(mas)} & \multicolumn{1}{c}{(km s$^{-1}$)} & \multicolumn{1}{c}{(km s$^{-1}$)} \\
\hline
\endhead
\hline
\endfoot 
    0 &        1,2,3,4 &       ... &    97.5 &  192.50$\pm$0.06  &   $-$399.23$\pm$0.07 & 0.0$\pm$0.0  & 0.0$\pm$0.0  \\
    1 &        1,2,3,4 &      9.57 &    92.3 &     0.00$\pm$0.00 &     0.00$\pm$0.00 &    2.8$\pm$1.1 &    3.4$\pm$1.2  \\
    2\tablefootmark{a} &        1,2,3,4 &      2.64 &   104.2 &   347.78$\pm$0.08 &   $-$47.91$\pm$0.09 &    ... &    ...  \\
    3 &        1,2,3,4 &      2.48 &   103.4 &   346.91$\pm$0.08 &   $-$38.57$\pm$0.08 &   $-$5.9$\pm$1.0 &   15.6$\pm$1.2  \\
    4 &        1,2,3,4 &      1.66 &    95.9 &  $-$241.15$\pm$0.09 &   $-$85.27$\pm$0.09 &   18.2$\pm$1.3 &   $-$3.0$\pm$1.3  \\
    5 &        1,2,3,4 &      1.63 &   103.6 &   356.28$\pm$0.08 &   $-$71.34$\pm$0.08 &  $-$18.9$\pm$1.2 &   18.7$\pm$1.4  \\
    6 &        1,2,3,4 &      1.14 &    93.8 &    25.76$\pm$0.08 &  $-$886.78$\pm$0.09 &    0.5$\pm$1.1 &   $-$8.1$\pm$1.3  \\
    7\tablefootmark{a} &        1,2,3,4 &      0.96 &    95.6 &  $-$254.83$\pm$0.09 &   $-$75.53$\pm$0.09 &    ... &    ...  \\
    8 &        1,2,3,4 &      0.95 &   102.9 &   357.14$\pm$0.08 &   $-$65.98$\pm$0.09 &   $-$8.9$\pm$1.1 &   16.9$\pm$1.4  \\
    9 &        1,2,3,4 &      0.86 &    96.0 &   596.01$\pm$0.09 &  $-$704.83$\pm$0.10 &   11.1$\pm$1.3 &  $-$11.5$\pm$1.6  \\
   10 &        1,2,3,4 &      0.84 &   105.4 &   340.86$\pm$0.08 &    $-$4.30$\pm$0.09 &   $-$2.2$\pm$1.4 &   10.7$\pm$1.7  \\
   11 &        1,2,3,4 &      0.80 &    95.6 &   591.89$\pm$0.09 &  $-$674.21$\pm$0.10 &   11.3$\pm$1.3 &  $-$10.1$\pm$1.5  \\
   12 &        1,2,3,4 &      0.77 &   102.7 &   353.48$\pm$0.08 &   $-$51.70$\pm$0.09 &   $-$7.3$\pm$1.2 &   14.8$\pm$1.5  \\
   13 &        1,2,3,4 &      0.75 &    96.3 &   543.18$\pm$0.09 & $-$1092.66$\pm$0.10 &    4.9$\pm$1.3 &   $-$7.6$\pm$1.6  \\
   14 &              1 &      0.75 &   103.5 &   349.46$\pm$0.10 &   $-$54.71$\pm$0.11 &    ... &    ...  \\
   15 &        1,2,3,4 &      0.74 &    95.2 &  $-$260.43$\pm$0.10 &   $-$72.04$\pm$0.10 &   14.5$\pm$1.3 &   $-$4.6$\pm$1.4  \\
   16 &        1,2,3,4 &      0.71 &    96.3 &  $-$134.94$\pm$0.11 &  $-$120.17$\pm$0.11 &   $-$4.8$\pm$1.8 &    6.2$\pm$1.9  \\
   17 &        1,2,3,4 &      0.61 &    98.8 &   624.01$\pm$0.09 &  $-$731.91$\pm$0.10 &    4.9$\pm$1.4 &   $-$4.7$\pm$1.7  \\
   18\tablefootmark{a} &        1,2,3,4 &      0.52 &    99.1 &   177.27$\pm$0.10 &   136.98$\pm$0.13 &    ... &    ...  \\
   19 &        1,2,3,4 &      0.42 &    99.2 &   171.80$\pm$0.12 &   147.41$\pm$0.14 &    1.5$\pm$1.7 &    6.0$\pm$2.2  \\
   20 &        1,2,3,4 &      0.35 &    93.3 &    55.71$\pm$0.11 &  $-$691.54$\pm$0.12 &    0.7$\pm$1.6 &   $-$9.9$\pm$2.0  \\
   21 &        1,2,3,4 &      0.32 &    96.8 &  $-$148.45$\pm$0.13 &  $-$114.68$\pm$0.13 &   $-$0.8$\pm$1.9 &    4.9$\pm$2.2  \\
   22\tablefootmark{a} &          1,2,3 &      0.31 &    98.7 &   616.63$\pm$0.10 &  $-$729.16$\pm$0.11 &    ... &    ...  \\
   23 &        1,2,3,4 &      0.29 &    97.6 &   604.58$\pm$0.12 &  $-$711.77$\pm$0.13 &   $-$1.9$\pm$1.8 &   $-$3.3$\pm$2.2  \\
   24 &        1,2,3,4 &      0.24 &   106.6 &   334.10$\pm$0.14 &   $-$20.74$\pm$0.18 &  $-$13.8$\pm$1.8 &   17.8$\pm$2.5  \\
   25 &        1,2,3,4 &      0.23 &    97.2 &    75.48$\pm$0.18 &   181.85$\pm$0.29 &    1.5$\pm$2.4 &    4.4$\pm$4.2  \\
   26 &        1,2,3,4 &      0.23 &   105.1 &   339.26$\pm$0.11 &   $-$68.06$\pm$0.13 &   $-$6.6$\pm$2.1 &   17.8$\pm$2.7  \\
   27 &        1,2,3,4 &      0.20 &    92.6 &    91.99$\pm$0.16 &  $-$683.14$\pm$0.18 &    7.6$\pm$2.5 &  $-$15.4$\pm$3.1  \\
   28 &            1,2 &      0.19 &   105.5 &   338.14$\pm$0.15 &   $-$70.04$\pm$0.19 &    ... &    ...  \\
   29 &            1,2 &      0.18 &    96.1 &  $-$105.40$\pm$0.19 &  $-$151.75$\pm$0.22 &    ... &    ...  \\
   30 &        1,2,3,4 &      0.17 &    98.2 &   118.15$\pm$0.24 &   116.94$\pm$0.38 &    3.3$\pm$3.1 &    3.8$\pm$4.7  \\
   31 &              4 &      0.17 &    98.5 &   614.03$\pm$0.18 &  $-$728.71$\pm$0.23 &    ... &    ...  \\
   32 &            1,2 &      0.17 &    96.2 &  $-$263.36$\pm$0.24 &   $-$78.72$\pm$0.21 &    ... &    ...  \\
   33 &          1,2,3 &      0.16 &    98.4 &   612.57$\pm$0.16 &  $-$727.02$\pm$0.17 &   $-$1.9$\pm$5.0 &   $-$8.6$\pm$5.8  \\
   34 &        1,2,3,4 &      0.16 &    94.9 &    $-$1.41$\pm$0.18 &  $-$989.38$\pm$0.17 &   $-$1.9$\pm$3.0 &   $-$5.9$\pm$3.3  \\
   35 &          1,2,4 &      0.16 &    95.4 &   577.99$\pm$0.27 &  $-$669.69$\pm$0.32 &   21.0$\pm$4.2 &  $-$10.1$\pm$4.8  \\
   36\tablefootmark{a} &        1,2,3,4 &      0.16 &    94.4 &    27.33$\pm$0.20 &  $-$990.87$\pm$0.26 &    ... &    ...  \\
   37 &        1,2,3,4 &      0.15 &    97.3 &  $-$133.29$\pm$0.24 &  $-$128.33$\pm$0.27 &   $-$6.7$\pm$3.4 &    9.0$\pm$4.0  \\
   38 &        1,2,3,4 &      0.15 &    92.6 &    57.45$\pm$0.21 &  $-$684.51$\pm$0.22 &    1.8$\pm$3.2 &  $-$14.1$\pm$3.7  \\
   39\tablefootmark{a} &          1,2,3 &      0.15 &    94.6 &    22.81$\pm$0.16 &  $-$983.89$\pm$0.20 &    ... &    ...  \\
   40 &        1,2,3,4 &      0.15 &    99.8 &   635.73$\pm$0.19 &  $-$746.47$\pm$0.25 &    1.3$\pm$3.1 &   $-$4.1$\pm$4.0  \\
   41 &        1,2,3,4 &      0.14 &   100.0 &   714.67$\pm$0.17 & $-$1004.55$\pm$0.19 &    5.4$\pm$3.9 &   $-$8.9$\pm$4.2  \\
   42\tablefootmark{a} &        1,2,3,4 &      0.14 &    93.0 &    36.56$\pm$0.30 &  $-$694.89$\pm$0.28 &    ... &    ...  \\
   43 &          1,2,3 &      0.13 &    95.2 &    14.57$\pm$0.25 &  $-$978.73$\pm$0.23 &   $-$0.4$\pm$8.2 &  $-$22.1$\pm$8.6  \\
   44 &            1,2 &      0.13 &    95.9 &   $-$95.07$\pm$0.28 &  $-$138.93$\pm$0.37 &    ... &    ...  \\
   45 &              1 &      0.13 &    95.8 &   587.12$\pm$0.28 &  $-$667.66$\pm$0.54 &    ... &    ...  \\
   46 &        1,2,3,4 &      0.11 &    94.7 &  $-$341.89$\pm$0.16 &  $-$474.74$\pm$0.20 &   $-$5.4$\pm$4.0 &    0.0$\pm$5.1  \\
   47 &        1,2,3,4 &      0.10 &    94.3 &  $-$113.82$\pm$0.22 &  $-$489.24$\pm$0.28 &   $-$0.3$\pm$3.7 &   $-$9.6$\pm$5.0  \\
   48 &        1,2,3,4 &      0.10 &    94.5 &    14.55$\pm$0.17 &  $-$989.79$\pm$0.20 &  $-$11.1$\pm$4.6 &   $-$8.0$\pm$5.4  \\
   49 &        1,2,3,4 &      0.10 &    92.0 &    57.76$\pm$0.25 &  $-$825.33$\pm$0.36 &    5.9$\pm$4.5 &  $-$14.3$\pm$6.5  \\
   50 &              2 &      0.09 &    93.8 &    18.28$\pm$0.25 &    41.24$\pm$0.41 &    ... &    ...  \\
   51 &              3 &      0.09 &    98.4 &   104.20$\pm$0.29 &   137.52$\pm$0.43 &    ... &    ...  \\
   52 &              2 &      0.09 &    94.5 &    14.85$\pm$0.29 & $-$1051.79$\pm$0.35 &    ... &    ...  \\
   53\tablefootmark{a} &        1,2,3,4 &      0.08 &    95.0 &  $-$270.45$\pm$0.40 &   $-$71.69$\pm$0.31 &    ... &    ...  \\
   54 &        1,2,3,4 &      0.08 &    93.0 &    47.88$\pm$0.18 &  $-$690.79$\pm$0.22 &    1.2$\pm$5.1 &  $-$10.7$\pm$6.0  \\
   55\tablefootmark{a} &        1,2,3,4 &      0.08 &    94.3 &    $-$9.62$\pm$0.39 &   100.93$\pm$0.62 &    ... &    ...  \\
   56 &        1,2,3,4 &      0.08 &    94.9 &   572.04$\pm$0.31 &  $-$658.57$\pm$0.43 &    4.3$\pm$5.2 &   $-$8.7$\pm$6.6  \\
   57\tablefootmark{a} &        1,2,3,4 &      0.08 &   106.7 &   320.07$\pm$0.19 &   $-$86.91$\pm$0.28 &    ... &    ...  \\
   58 &              1 &      0.08 &    94.9 &     8.04$\pm$0.39 &  $-$988.51$\pm$0.47 &    ... &    ...  \\
   59 &            1,2 &      0.08 &    93.9 &   134.44$\pm$0.23 &  $-$730.51$\pm$0.35 &    ... &    ...  \\
   60 &        1,2,3,4 &      0.08 &    97.9 &    86.04$\pm$0.38 &   152.20$\pm$0.57 &    2.2$\pm$5.1 &   11.6$\pm$8.0  \\
   61 &          1,2,3 &      0.07 &   101.0 &   727.27$\pm$0.31 &  $-$917.53$\pm$0.38 &   13.4$\pm$9.0 &  $-$19.4$\pm$11.1  \\
   62 &          1,2,3 &      0.07 &    97.0 &    53.64$\pm$0.36 &   177.40$\pm$0.59 &    1.2$\pm$8.2 &    5.0$\pm$13.2  \\
   63 &            1,2 &      0.07 &   104.5 &   301.71$\pm$0.25 &   $-$80.64$\pm$0.42 &    ... &    ...  \\
   64 &              1 &      0.07 &    96.5 &   $-$48.64$\pm$0.35 &  $-$844.58$\pm$0.48 &    ... &    ...  \\
   65 &              1 &      0.06 &    94.5 &  $-$122.40$\pm$0.39 &  $-$492.51$\pm$0.43 &    ... &    ...  \\
   66 &              1 &      0.06 &    93.2 &   604.31$\pm$0.37 &  $-$664.64$\pm$0.48 &    ... &    ...  \\
   67 &              2 &      0.06 &    98.6 &   103.39$\pm$0.62 &   142.21$\pm$1.28 &    ... &    ...  \\
   68 &              4 &      0.06 &    98.5 &   103.72$\pm$0.76 &   139.63$\pm$1.28 &    ... &    ...  \\
   69 &              1 &      0.06 &    95.0 &    98.92$\pm$0.40 &  $-$822.30$\pm$0.52 &    ... &    ...  \\
   70\tablefootmark{a} &          1,3,4 &      0.06 &    97.0 &  $-$239.70$\pm$0.49 &   $-$82.07$\pm$0.68 &    ... &    ...  \\
   71 &              1 &      0.05 &    98.6 &   102.52$\pm$0.47 &   144.12$\pm$0.88 &    ... &    ...  \\
   72 &            1,2 &      0.05 &    93.6 &   $-$17.51$\pm$0.39 &    38.20$\pm$0.64 &    ... &    ...  \\
   73 &              1 &      0.05 &    93.8 &   115.21$\pm$0.35 & $-$1317.51$\pm$0.56 &    ... &    ...  \\
   74 &              1 &      0.05 &   106.6 &   340.24$\pm$0.29 &    $-$5.02$\pm$0.45 &    ... &    ...  \\
   75 &            1,2 &      0.05 &    93.1 &    98.92$\pm$0.69 &  $-$676.75$\pm$0.64 &    ... &    ...  \\
   76 &            1,2 &      0.05 &    93.6 &    $-$8.51$\pm$0.39 &    13.39$\pm$0.72 &    ... &    ...  \\
   77\tablefootmark{a} &          1,2,4 &      0.05 &    98.0 &   527.33$\pm$0.40 &  $-$726.68$\pm$0.66 &    ... &    ...  \\
   78 &              1 &      0.05 &    94.2 &    $-$6.38$\pm$0.46 &   204.54$\pm$0.63 &    ... &    ...  \\
   79 &              2 &      0.05 &    91.5 &    33.46$\pm$0.38 &    56.86$\pm$0.66 &    ... &    ...  \\
   80 &              1 &      0.05 &    92.7 &    83.34$\pm$0.59 &  $-$690.23$\pm$0.80 &    ... &    ...  \\
   81 &              3 &      0.05 &   105.8 &   351.51$\pm$0.53 &  $-$160.45$\pm$0.67 &    ... &    ...  \\
   82 &              1 &      0.04 &    94.3 &    $-$9.02$\pm$0.48 &   213.87$\pm$0.73 &    ... &    ...  \\
   83 &              3 &      0.04 &    94.7 &   $-$46.01$\pm$0.71 &   146.18$\pm$0.84 &    ... &    ...  \\
   84 &              1 &      0.04 &    90.5 &  $-$137.65$\pm$0.46 &  $-$532.52$\pm$0.77 &    ... &    ...  \\
   85 &              1 &      0.04 &    97.0 &  $-$239.12$\pm$0.66 &   $-$91.77$\pm$0.63 &    ... &    ...  \\
\end{longtable}
\tablefoot{Column~1 gives the feature label number; column~2
lists the observing epochs at which the feature was detected;
columns~3~and~4 provide the intensity of the strongest spot
and the intensity-weighted LSR velocity, respectively, averaged over the
observing epochs; columns~5~and~6 give the position offsets (with
the associated errors) along the R.A. and Dec. axes, relative to the
feature~\#1, measured at the first epoch of detection; columns~7~and~8 give the components of the relative
proper motion (with the associated errors) along the R.A. and Dec.
axes, measured with respect to the reference feature~\#0 (the
``center of motion'').\\
\tablefoottext{a}{Feature with not reliable proper motion, even if observed at three or four epochs.}
}
}

\end{document}